\documentclass[prd,singlecolumn,superscriptaddress,nofootinbib,10pt]{revtex4}
\usepackage{graphicx,amssymb,amsmath,slashed}
\usepackage{color,graphicx,epsfig}
\usepackage{ifpdf}
\usepackage{amsfonts}

\usepackage{braket}

\definecolor{nicered}{rgb}{0.7,0.1,0.1}
\definecolor{nicegreen}{rgb}{0.1,0.5,0.1}

\newcommand{\DS}[1]{/\!\!\!#1}
\newcommand{\beq}{\begin{equation}}
\newcommand{\eeq}{\end{equation}}

\begin{document}

\title{Discerning New Physics in $t \bar t$ Production using Top Spin Observables at  Hadron Colliders}

\author{Svjetlana Fajfer} 
\affiliation{Department of Physics,
  University of Ljubljana, Jadranska 19, 1000 Ljubljana, Slovenia}
\affiliation{J. Stefan Institute, Jamova 39, P. O. Box 3000, 1001 Ljubljana, Slovenia}

\author{Jernej F. Kamenik} 
\affiliation{Department of Physics,
  University of Ljubljana, Jadranska 19, 1000 Ljubljana, Slovenia}
\affiliation{J. Stefan Institute, Jamova 39, P. O. Box 3000, 1001 Ljubljana, Slovenia}

\author{Bla\v zenka Meli\' c }
\affiliation{Rudjer Bo\v skovi\' c Institute, Theoretical Physics Division,Bijeni\v cka c. 54, 10000 Zagreb, Croatia}

\date{\today}

\begin{abstract}
Copious production of top - anti top quark pairs at hadron colliders has enabled various probes into the properties and interactions of top quarks. Among the various presently measured observables, the forward-backward asymmetry (FBA) in $t\bar t$ production measured at the Tevatron significantly deviates from the standard model predictions, and many models of new physics have been invented to explain the puzzle. 
We consider the consistency of the simplified single-resonance models containing a color octet axial-vector (``axigluon"), color triplet or sextet weak singlet scalars, weak isodoublet scalar, flavor-changing neutral $Z'$, or charged $W'$ vector boson with existing $t\bar t$ production measurements. Among the considered models only an axigluon can reproduce all Tevatron observables, without  being in severe tension with the recent LHC results on $t\bar t$ production cross section, charge asymmetry and  top-spin correlations. The LHC charge asymmetry measurements exclude the $W'$ and $Z'$ explanations of the Tevatron FBA anomaly. On the other hand, all scalar models predict notable deviations in several top spin observables, and the recent top spin correlation measurement using the ``helicity" spin quantization axis by ATLAS already provides a significant constraint on possible explanations of the Tevatron FBA anomaly. Future precise measurements of top spin correlations and especially top polarization could differentiate between scalar $t$-channel models, while they are less sensitive to pure axigluon contributions.
\end{abstract}

\maketitle
 
%
\section{INTRODUCTION}
%

Excellent performances of the Tevatron and recently the LHC have enriched our knowledge of top quark physics with many interesting results. Most intriguing are the observed deviations from the Standard Model (SM) predictions for the Forward-Backward Asymmetry (FBA) in $t \bar t$ production at Tevatron. 
At the same time the measured inclusive $t \bar t$  production cross section and its $t\bar t$ invariant mass  ($m_{t\bar t}$) distribution are in a good agreement with SM predictions. Similarly, early $t\bar t$ production Charge Asymmetry (CA) measurements at the LHC show no deviations from zero, consistent with tiny SM predictions.
This observed pattern of deviations motivated many  studies of New Physics (NP)  aiming to explain the puzzling phenomena. 
The required size of possible NP contributions to the FBA points towards tree-level effects and most studies employ the  description  in terms of $s$, $t$ or $u$-channel resonance exchanges (c.f.~\cite{ttreview} for a recent review).  There are  several candidate states which can serve to explain the Tevatron experimental results, such as the  $s$-channel axial-vector color octet (``axigluon"),  or alternatively $u$, or $t$-channel exchanged scalars (isosinglet color triplet or sextet, and color-neutral weak isodoublet) or flavor-changing neutral ($Z^\prime$) or charged ($W'$) vector bosons.

In order to shed light on the possible role of NP in top - anti top  pair production at the Tevatron and the LHC,  it is important to investigate all possible observables. For more than a decade proposals have been made to investigate top quark spin polarization and correlation effects. Namely, due to its very short lifetime, top quark spin information is not spoilt by hadronization effects and can be reconstructed from the angular distribution of decaying particles~\cite{ttspin}. 
With the large $t\bar t$ datasets at the LHC, the possible presence of NP in $t\bar t$ production might be tested through new variables such as the (anti)top spin polarization and top-antitop spin-spin correlations~\cite{Krohn:2011tw, Cao:2011hr, Bai:2011uk}. The intricate relations between the $t\bar t$ charge asymmetric production, the underlying helicity amplitude structure and angular distributions of top quark decay products in $t\bar t$ production have also been pointed out recently in~\cite{Falkowski:2011zr, Berger:2012nw}.

Motivated by these developments and recent first precision measurements of CA and top spin observables at the LHC we reinvestigate effective single resonance NP models which aim to address the Tevatron anomalous FBA. Performing a fit of NP model parameters taking into account the presently measured $t\bar t$ production observables at the Tevatron and the LHC we determine the viable parameter regions in each model (previously, a similar analysis using the first LHC data has been performed in~\cite{AguilarSaavedra:2011ug}). In this we pay special attention to the apparent tension between the large FBA values as measured at the Tevatron and the existing tight constraints on the CA from the LHC. For the viable model parameter regions we predict top spin observables at the Tevatron and the LHC, evaluate the impact of existing measurements, and point out which near future experimental analyses could discriminate between the various NP proposals for the FBA puzzle.

This paper is organized as follows: in Secs.~\ref{sec:models} and~\ref{sec:obs} we introduce the effective NP resonance models addressing the FBA anomaly, and the relevant Tevatron and LHC observables constraining these models, respectively. The concordance of the NP models with existing $t\bar t$ production measurements is quantified in Sec.~\ref{sec:fit} and the corresponding best fit regions are identified. In Sec.~\ref{sec:Spin} we discuss the top spin phenomenology in hadronic $t\bar t$ production and give the FBA correlated NP model predictions for the relevant top spin observables at the Tevatron and the LHC. Finally we conclude in Sec.~\ref{sec: conclusions}

%
\section{Models}
\label{sec:models}
%
We consider NP models which affect the $t\bar t$ production at hadronic colliders at the tree level and consider a single NP amplitude interfering with the SM contributions at a time. Such scenarios can then be classified according to the new resonances coupling to quarks and exchanged in $s-$, $t-$ or $u-$channel. 
We note however that our analysis is applicable also to NP models with more new degrees of freedom accessible at low energies (as for example in approximately flavor symmetric models~\cite{JureBenAlex}) but where the $t\bar t$ production phenomenology is dominated by the exchanges of a single intermediate state. Among the plethora of possible spin, weak isospin, charge and color assignments, only a few of such states can produce a sizable positive FBA at the Tevatron without being in gross conflict with the measurements of the total cross-section and/or the $m_{t\bar t}$ spectrum~\cite{AguilarSaavedra:2011vw}. These include an $s-$channel exchanged (axial-)vector color octet (axigluon $G'$) boson~\cite{axig,Barcelo,Schmalz}, neutral ($Z'$)~\cite{Zp} or charged ($W'$)~\cite{Wp} vector bosons, coupling chirally to quarks and exchanged in the $t-$channel, a scalar isodoublet~\cite{doublet} whose neutral component ($\phi^0$) contributes in the $t-$channel,  as well as scalar color triplet ($\Delta$)~\cite{Shu:2009xf,triplet} or sextet ($\Sigma$)~\cite{Shu:2009xf} coupling chirally to up-type quarks and contributing in the $u-$channel. The relevant interaction Lagrangians for all the considered models can be found in Appendix A.

%
\section{Observables \& Methodology}
\label{sec:obs}
%
We probe the parameter space of considered NP models with presently measured observables in $t\bar t$ production both at the Tevatron and the LHC. 
First, there is the inclusive FBA as measured by both the CDF~\cite{AFBCDF} and D\O~\cite{AFBD0} collaborations. Combining all the measurements on independent datasets by both experiments and adding the uncertainties in quadrature leads to a na\"ive average
\beq
A_{FB} = 0.187\pm 0.037\,,
\eeq
compared to the NLO QCD prediction~\cite{AFBNLO,AFBCDF,MCNLO} (consistent with recent NLO+NNLL estimates~\cite{AFBSM} within errors) including leading
electroweak (EW) contributions~\cite{AFBEW,Manohar:2012rs} of $A^{\rm SM}_{FB} = 0.07(2)$. At the LHC, the most precise measurements of the Charge Asymmetry (CA) to date  by  ATLAS~\cite{ACATLAS} and CMS~\cite{ACCMS} can again be na\"ively combined to yield
\beq
A_C = 0.001 \pm 0.014\,,
\eeq
and compared to the SM prediction of $A_C^{\rm SM} = 0.007(1)$~\cite{AFBNLO,ACATLAS,MCNLO,AFBEW,Manohar:2012rs}. 

On the other hand, while both CDF and D\O \, have also presented FBA measurements in bins of $m_{t\bar t}$ and top - anti top rapidity differences, only CDF~\cite{AFBCDF} presents results unfolded to the partonic (truth) level with the result
for two $m_{t\bar t}$ bins
\begin{subequations}
\begin{eqnarray}
A^l_{FB}\equiv A_{FB}(m_{t\bar t}<450~{\rm GeV}) &=& 0.078 \pm 0.054\,,\\
A^h_{FB} \equiv A_{FB}(m_{t\bar t}>450~{\rm GeV}) &=& 0.296 \pm 0.067\,,
\end{eqnarray}
\end{subequations}
to be compared with SM predictions~\cite{AFBNLO,AFBCDF,MCNLO,AFBEW,Manohar:2012rs} of $A^{l,\rm SM}_{FB} = 0.05(1)$ and $A^{h,\rm SM}_{FB} = 0.11(2)$. Since these measurements are not independent of the inclusive FBA average,  we take two complementary approaches. In the first (approach A) we use the inclusive $A_{FB}$, while in the second (B) we instead consider only the two binned FBA values $A^{l,h}_{FB}$ (we take the associated uncertainties as uncorrelated). In both approaches we also treat the production cross-section measurements a bit differently (see below). We compare the constraints on NP model parameters in both approaches (A and B) in the next section.
Similar $m_{t\bar t}$ binned CA measurements have been performed by ATLAS~\cite{ATLASACmtt} and CMS~\cite{ACCMS}, but with large uncertainties, which make these observables at present less constraining than the inclusive CA measurements. 

Also crucially important observables at the Tevatron are the total inclusive $t\bar t$ production cross-section, with a recent experimental average of~\cite{sigmaTEV}
\beq
\sigma_{\rm TEV} = (7.5\pm 0.48) \,\rm pb\,,
\label{eq:sigmaTEV}
\eeq
compared to the recent NNLO+NNLL QCD prediction of $\sigma_{\rm TEV}^{\rm SM} = 7.07(26)~\,\rm pb$~\cite{sigmaTEVSM}; as well as the $m_{t\bar t}$ spectrum  ($d\sigma_{\rm TEV}/dm_{t\bar t}$) measured in several bins by the CDF collaboration~\cite{CDFmtt} and in good agreement with NLO+NNLL QCD predictions~\cite{mttSM}.\footnote{It has been pointed out recently, that sizable EW corrections can {\it reduce} the SM $d\sigma_{}/dm_{t\bar t}$ predictions in the high $m_{t\bar t}$ region~\cite{Manohar:2012rs}, thus leaving more room for potential NP contributions. However, such effects are smaller than the current experimental uncertainties~\cite{CDFmtt} and can at present be safely neglected.} Since again the two observables are not independent, we consider one of them in each of the two complementary approaches (A and B) discussed above. In the first (approach A) we use $\sigma_{\rm TEV}$ but also the next-to-last bin ($m_{t\bar t} \in[700,800]$~GeV) of the CDF $m_{t\bar t}$ spectrum measurement~\cite{CDFmtt}
\beq
\sigma^h_{\rm TEV}  = (80\pm 37) \,\rm fb\,,
\eeq
to be compared with the SM expectation of $\sigma^{h,\rm SM}_{\rm TEV}  = 80(8) \,\rm fb$~\cite{mttSM}. While the two observables are not completely independent, the effect of the high $m_{t\bar t}$ tail of the spectrum on the total cross-section measurement is negligible (i..e much smaller than the experimental error of $\sigma_{\rm TEV}$). In the second approach (B) we do not use the total inclusive cross-section $\sigma_{\rm TEV}$ at all, but instead employ all the $m_{t\bar t}$ bins of the measured CDF spectrum. In this we consider the statistical uncertainties  as uncorrelated among the bins, but we use a 100\% correlation approximation for both the systematic as well as SM theoretical uncertainties.   

At the LHC, the most precise inclusive $t\bar t$ production cross-section determinations are presently provided by ATLAS~\cite{sigmaATLAS} and CMS~\cite{sigmaCMS} when combining their dilepton, single-lepton and all-hadronic final state analyses. Again performing a na\"ive average of both experiments, we obtain
\beq
\sigma_{\rm LHC} = (172\pm 10)\, \rm pb\,, 
\eeq
in good agreement with the approximate NNLO QCD prediction of $\sigma^{\rm SM}_{\rm LHC} = (163^{+11}_{-10})$\,pb~\cite{sigmaLHCSM}.  On the other hand the existing high $m_{t\bar t}$ spectrum measurements~\cite{mttLHC} are limited by moderate statistics and significant systematic uncertainties~\cite{topneutron}. In particular, in the $m_{t\bar t} > 1$~TeV region they yield the constraint~\cite{mttLHC}
\beq
\frac{\sigma^h_{\rm LHC}}{\sigma^{h,\rm SM}_{\rm LHC}} < 2.6 ~ @\, 95\%\, \rm C.L.\,.
\eeq 
Since the dominant sources of uncertainty should be significantly reduced with more data, the $m_{t\bar t}$ spectrum measurements could yield important constraints on NP contributions to $t\bar t$ production in the near future. 

Finally, $t\bar t$ spin correlations (as defined in Sec.~\ref{sec:Spin:Def}) have been observed both at the Tevatron~\cite{spinTEV,spinTEV1} and the LHC~\cite{spinLHC,Aad:2012sm}. At the Tevatron presently the most precise determination has been performed in the ``beamline" basis by the D\O~collaboration~\cite{spinTEV1}, obtaining 
\beq
C^{\rm TEV}_{\rm beam} = 0.66 \pm 0.23\,,
\eeq
in agreement with NLO QCD prediction of  $C^{\rm TEV}_{\rm beam,SM} =  0.78^{+0.03}_{-0.04}$~\cite{spinSM}. On the other hand, the most recent ``helicity" basis analysis by CDF~\cite{spinTEV} yields
\beq
C^{\rm TEV}_{\rm hel} = -0.60 \pm 0.52\,,
\eeq
where we have combined the statistical and systematic uncertainties in quadrature. Within the large uncertainties this is in agreement with the SM expectation of $C^{\rm TEV}_{\rm hel, SM} \simeq -0.35$~\cite{spinSM}. At the LHC, the recent ATLAS measurement~\cite{Aad:2012sm} of the ``helicity" axis spin correlation coefficient 
\beq
C^{\rm LHC}_{\rm hel} = 0.40^{+0.09}_{-0.08}\,,
\eeq
in agreement with the SM prediction of $C_{\rm hel}^{\rm LHC,SM} \simeq 0.31$~\cite{spinSM} already provides a significant constraint on possible explanations of the Tevatron FBA anomaly. A detailed discussion of the impact of present and future measurements of $t\bar t$ spin observables on the NP models under consideration is given in section~\ref{sec:Spin}.

Some of the models we consider can also be constrained by phenomenology not directly related to $t\bar t$ production like same-sign and single top production, dijets, electroweak precision observables and rare B processes (c.f.~\cite{ttreview} for a recent review and also~\cite{Gresham:2012wc}). Since in the present analysis we want to focus on the discriminating power of $t\bar t$ observables at the Tevatron and the LHC, we will assume that other constraints can be evaded in suitable UV completions of the considered effective single-resonance models.

We compute the SM+NP contributions to the selected observables at the partonic level employing the MSTW2008 parton distribution functions~\cite{MSTW} at fixed factorization and renormalization scale $\mu_F=\mu_R=m_t=172.5$~GeV -- the top quark mass used also in the above quoted measurements to which we compare our predictions. For all observables we compute all interfering SM (QCD) and NP contributions at leading order in $\alpha_s$ (at the tree level).  We then normalize our SM+NP tree-level estimates to the pure SM result, i.e. we define normalized cross-sections $\bar \sigma$
\begin{equation}
\bar \sigma^{\rm th} = \frac{\sigma^{\rm LO}}{\sigma^{\rm LO}_{\rm QCD}}\,,
\label{eq:sigmaExp}
\end{equation}
and compare these to the experimental cross-sections normalized to the state-of-the-art SM predictions,
\begin{equation}
\bar \sigma^{\rm exp} = \frac{\sigma^{\rm exp}}{\sigma_{\rm SM}}\,.
\end{equation}
In this way we hope to capture a (universal) part of the higher order QCD corrections to the observables under study.~\footnote{Presently, leading QCD corrections have only been computed within the $W'$~\cite{QCDWp} and axigluon~\cite{QCDAG} NP models and for a limited set of observables. These known results suggest that higher order QCD effects in NP contributions are indeed similar in size to SM higher order QCD corrections.} 
In the case of forward-backward, charge asymmetries and top spin observables which are defined normalized to the production cross section, we first subtract our pure LO QCD estimates to obtain an estimate of the NP (and interference) contributions, i.e. for observable $\mathcal O$ we define
\begin{equation}
\Delta \mathcal O^{\rm th} \equiv \mathcal O^{\rm LO} - \mathcal O^{\rm LO}_{\rm QCD}\,.
\end{equation}
 We compare these to the experimental values for the asymmetries and top spin observables from which we have subtracted the state-of-the-art SM predictions, corrected for the change in the normalized cross-section
\begin{equation}
\Delta \mathcal O^{\rm exp} \equiv \mathcal O^{\rm exp} - \frac{1}{\bar \sigma^{\rm th}}\mathcal O_{\rm SM} \,.
\end{equation} 
In practice the strong experimental bound on NP contributions to $\sigma_{\rm TEV}$ in Eq.~\eqref{eq:sigmaTEV} restricts $\bar \sigma^{\rm th}\simeq 1$ for inclusive observables (especially at the LHC due to the dominance of the gluon-fusion subprocess, not affected by NP in all models under consideration) and the associated corrections to $\Delta \mathcal O^{\rm exp}$ are always small.

A second issue regards experimental acceptance corrections due to limited rapidity coverage of the CDF and D\O\, detectors. In deconvolving the measured detector level observables to the truth (partonic) level, the existing analyses assume (anti)top rapidity distributions resembling the SM predictions. However, NP models with light mediators exchanged in the $t-$ and $u-$ channels can exhibit a forward scattering peak thus enhancing (or suppressing) the cross-section relative to the SM at high rapidities~\cite{Gresham}. To correct for these effects we employ the procedure suggested in~\cite{JureBenAlex} by computing and applying efficiency corrections to the NP cross-section distributions in bins of top - anti top rapidity difference and $m_{t\bar t}$. 

In addition to the above mentioned acceptance effects, $t-$ or $u-$ channel models can contribute to $t\bar t +$jet production for resonance masses above $m_t$ via associated resonance-top production. This could in principle affect also the $t\bar t $ measurements by increasing the reconstructed cross-sections~\cite{Gresham}. 
The magnitude of such effects however is proportional to the branching fraction for the decay of the resonances into $t+$jet which could in principle be reduced in UV completions of the effective models by the presence of other decay channels.  In addition their relevance also depends somewhat on the details of the experimental analyses.  Consequently we do not include them in our procedure. 

Finally, $t-$ or $u-$ channel resonances with masses below $m_t$ can contribute new decay channels of the top quark, thus enhancing its width. While such effects are constrained by the single top production measurements~\cite{singleTop}, the exact sensitivity of existing measurements is difficult to asses. Thus we do not strictly impose any such constraint, but do highlight such effects if significant in certain part of NP model parameter space. On the other hand, all of the present $t\bar t$ production experimental analyses rely on the kinematic reconstruction of top quarks assuming the standard decay mode $t\to b W$. In particular they require tight reconstruction criteria for the $W$ in both hadronic and leptonic final states (c.f.~\cite{CDFPhD}). We will assume that any new decay mode of the top quark would not pass the top reconstruction criteria thus effectively reducing the measured cross-section. We take this effect into account by multiplying the predicted cross-sections by the square of the ratio of the total SM top width (assuming $V_{tb}=1$) and the NP enhanced top width (both computed at leading order in QCD). In particular, we correct Eq.~\eqref{eq:sigmaExp} to
\begin{equation}
\bar \sigma^{\rm th} = \frac{\sigma^{\rm LO}}{\sigma^{\rm LO}_{\rm QCD}} {\rm Br}(t\to b W)^2\,.
\label{eq:Br}
\end{equation}

%
\section{ Model Fits }
\label{sec:fit}   
%

In constraining the NP models from $t\bar t$ production phenomenology we combine the observables discussed in the previous section into a global $\chi^2$ fit ($A_{FB},\,A_C,\, \sigma_{\rm TEV}$, $\sigma_{\rm TEV}^h$, $\sigma_{\rm LHC}$ and $\sigma^h_{\rm LHC}$ in approach A, or $A^{l,h}_{FB},\,A_C$, $\sigma_{\rm LHC}$, $\sigma^h_{\rm LHC}$ and $d\sigma_{\rm TEV}/dm_{t\bar t}$ in approach B as explained in the previous section). While we do not assign a statistical significance to the particular $\chi^2$ values, we consider NP model parameter regions as acceptable if they improve upon the $\chi^2$ obtained in the pure SM ($\chi^2_{A,\rm SM} = 1.9$/d.o.f. in approach A and $\chi^2_{B,\rm SM} = 1.6$/d.o.f. in approach B). 
For these regions, which represent scenarios addressing the FBA discrepancy and being in reasonable agreement with other existing $t\bar t$ production measurements, we predict the relevant top spin observables at the Tevatron and the LHC in Sec.~\ref{sec:Spin}.

\begin{figure}[!t]
  \centering
  \includegraphics[width=0.47\textwidth]{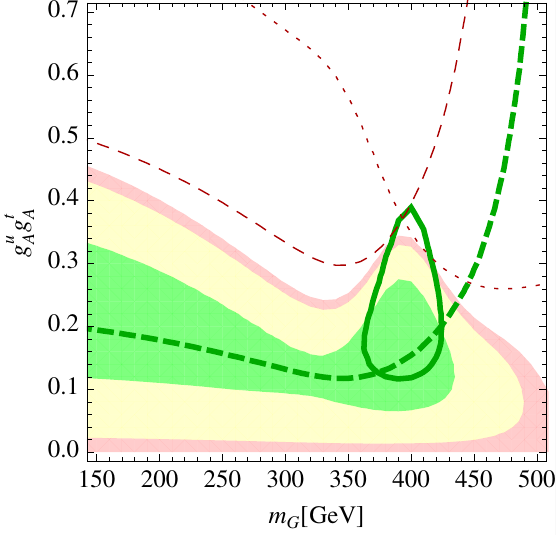}
  \includegraphics[width=0.47\textwidth]{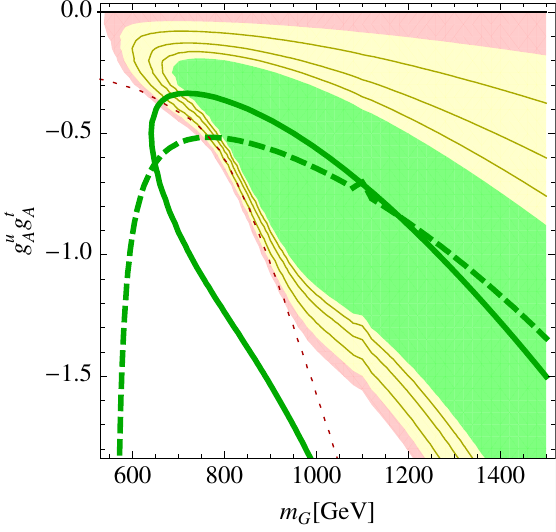}
  \caption{$t\bar t$ production constraints on the axigluon model in approach B, showing ``low" (left plot) and ``high" (right plot) mass regions: binned FBA at $1\sigma$ in thick full green line, inclusive CA at $1\sigma$ ($2\sigma$) in thick dashed green line (thin dashed red line), CDF $m_{t\bar t}$ spectrum at $2\sigma$ in thin red dotted line.  Parameter regions where the model can improve the SM $\chi^2$ by $-\Delta \chi^2 > 0,1,4$ are shaded in red, yellow and green respectively.} 
  \label{fig:axigluon_bounds}
\end{figure}

\subsection{Axigluon}

Among the possible interactions of a heavy color-octet (axial)vector $G'$, only the purely axial couplings to $u\bar u$ ($g_A^{u}$) and $t\bar t$ ($g_A^t$) currents contribute to the FBA at the tree-level.\footnote{Maintaining purely axial quark-axigluon couplings in an electroweak gauge invariant way requires identical couplings to down quarks. However, since at the Tevatron and the LHC, the corresponding $d\bar d$ parton lumionsities are much smaller than $u\bar u$,  we neglect the small $d\bar d$ initial state axigluon contributions to our observables.}  We thus consider their product $g_A^u g_A^t$ together with the axigluon mass ($m_G$) as the free parameters of the model. Since the effects of a $s-$channel resonance on the $m_{t\bar t}$ spectrum depend crucially on its width we follow the approach of~\cite{Barcelo,Schmalz} and assign an axigluon width of $\Gamma_G/m_G\sim 0.2$, deferring the explanation of such a large width to a UV completion of the model. The singular features of an $s-$channel resonance on $t\bar t$ spectrum also make the approach B (using the $m_{t\bar t}$ spectrum instead of the total cross-section and inclusive FBA) much preferred for the axigluon model.  The results of the global $\chi^2 $ fit together with the impact of individual constraints is presented in Fig.~\ref{fig:axigluon_bounds}. 
We single out two interesting parameter plane regions as the ``low" and ``high" mass region focused on the left and the right plots respectively (consistent with results of previous studies, c.f.~\cite{Bai:2011ed}). We observe that even with the large unexplained axigluon width, axigluon masses which fall within the measured CDF $m_{t\bar t}$ spectrum bins are significantly constrained and in particular cannot improve upon the SM agreement with current experimental results. On the other hand the finite axigluon width effects become insignificant in the ``low"  region around and below $t\bar t$ production threshold  (for $m_G\lesssim 450$~GeV, consistent with the findings of~\cite{Schmalz}) and in the ``high" region above the kinematic reach of the Tevatron (for $m_G>700$~GeV, as pointed out in~\cite{Barcelo}). At least in the later, $\sigma^h_{\rm LHC}$~\cite{mttLHC} already contributes a relevant constraint for $m_G\gtrsim 1$~TeV and future precise LHC $t\bar t$ spectrum measurements could have a significant impact. We also note that for this model the tension between FBA and CA measurements is apparent but can be reduced to the $1\sigma$ level. The tension is expected to be larger when symmetric couplings to $d$ quarks (as formally required by SM gauge invariance for a pure axigluon) are included~\cite{AFBvsAC}. 

\subsection{$Z'$ and $W'$}

\begin{figure}[!t]
  \centering
  \includegraphics[width=0.47\textwidth]{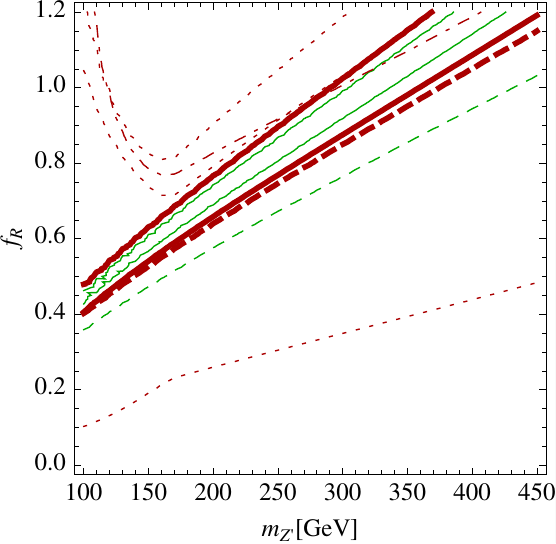}
  \includegraphics[width=0.47\textwidth]{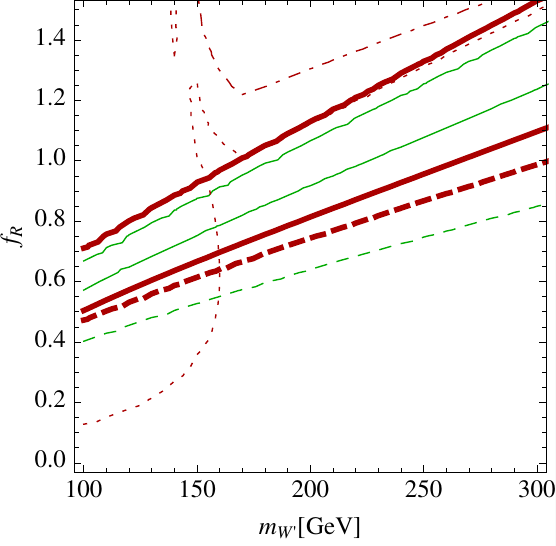}
  \caption{$t\bar t$ production constraints on the $Z'$ (left plot) and $W'$ (right plot) models  in approach A: FBA at $1\sigma$ ($2\sigma$)in thin full green line (thick full red line), CA at $1\sigma$ ($2\sigma$) in thin dashed green line (thick dashed red line), $\sigma_{\rm TEV}$ at $2\sigma$ in thin red dotted line, $\sigma^h_{\rm TEV}$ at $2\sigma$ in thin red dash-dotted line. In no parameter region can either of the two models improve upon the SM fit due to the tension between the FBA and CA measurements.} 
  \label{fig:Zp_bounds}
\end{figure}

\begin{figure}[!t]
  \centering
  \includegraphics[width=0.47\textwidth]{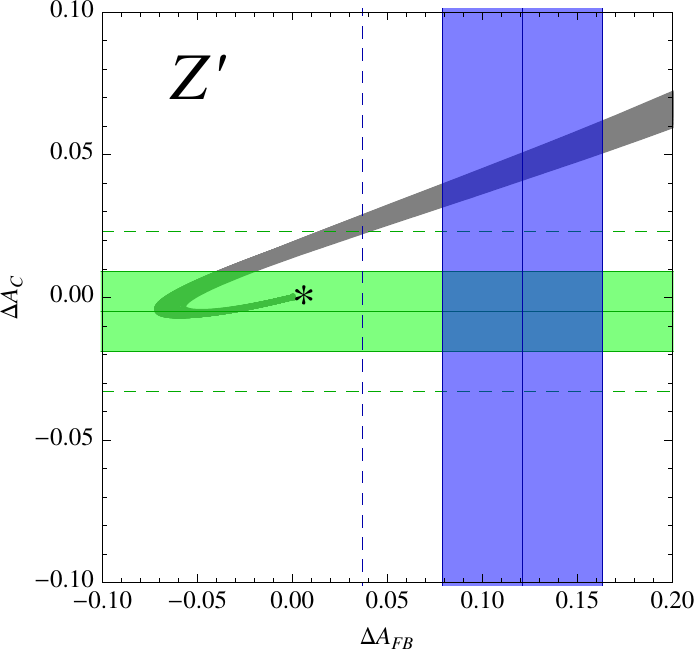}
  \includegraphics[width=0.47\textwidth]{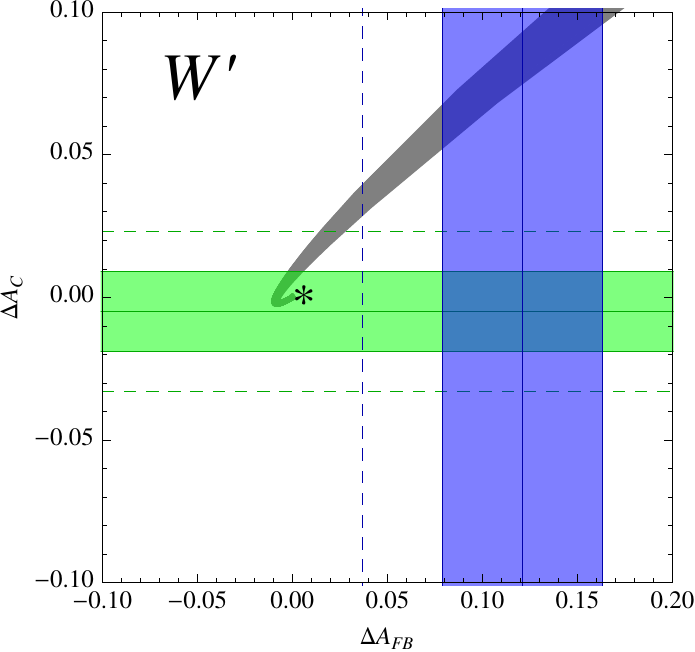}
  \caption{Correlation between the FBA ($\Delta A_{FB}$) and CA ($\Delta A_{C}$) contributions of $Z'$ (left plot) and $W'$ (right plot) models in shaded narrow gray bands. The thickness of the bands is given by the $Z'$ ($W'$) mass variation in the range $[100,500]$~GeV. The Tevatron average of FBA measurements minus the SM prediction at $1\sigma$ ($2\sigma$) is represented by the vertical blue band (dashed vertical lines). The ATLAS measurement of CA minus the SM prediction (the SM reference point is marked with "*") at $1\sigma$ ($2\sigma$) is represented by the horizontal green band (dashed horizontal lines). The tension between the two observables increases with increasing mediator mass (inner edge of the band corresponds to lowest mediator mass).} 
  \label{fig:ACvsAFB}
\end{figure}

For the $Z'$ and $W'$ models we consider only right-handed flavor changing $u-t$ and $d-t$ couplings (denoted by $f_R$) respectively since these are sufficient to induce a positive FBA contribution. The resulting parameter fits for approach A are represented in Fig.~\ref{fig:Zp_bounds}. 
We observe immediately that while a consistent description of $t\bar t$ observables at the Tevatron can be obtained (in accordance with previous findings~\cite{Bhattacherjee:2011nr}), these models cannot improve upon the SM results once the LHC measurements are taken into account. This can be traced to the inherent tension between the FBA and CA measurements (and thus independent of approach A or B to the cross-section measurements). For $Z'$($W'$) these are tightly correlated (as previously pointed out in~\cite{AguilarSaavedra:2011vw}), as also illustrated in Fig.~\ref{fig:ACvsAFB} where we have varied $f_R$ within the perturbative regime ($|f_R| \in [0,4\pi]$) and the $Z'$ ($W'$) masses in the region $m_{Z'(W')} \in[100,500]$~GeV. 
In particular from these plots it is clear that the tension cannot be reduced below the $2\sigma$ level. We also note that the tension increases slightly towards higher $Z'$ ($W'$) masses (when marginalized over the couplings). Consequently, we conclude that these models are excluded as viable FBA explanations and do not consider them any further.

\subsection{Scalar isodoublet}

\begin{figure}[!t]
  \centering
  \includegraphics[width=0.47\textwidth]{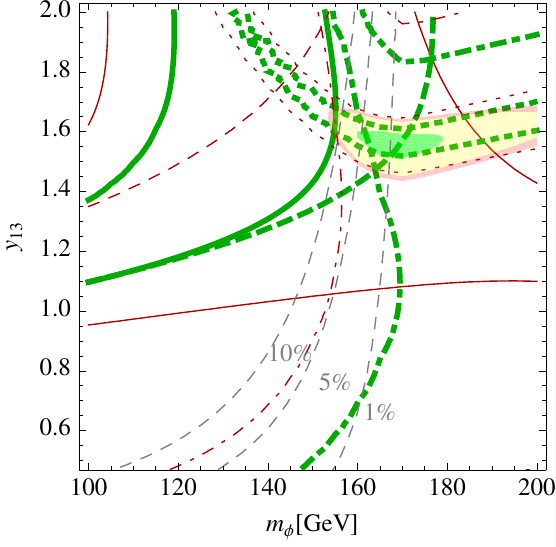}
  \includegraphics[width=0.47\textwidth]{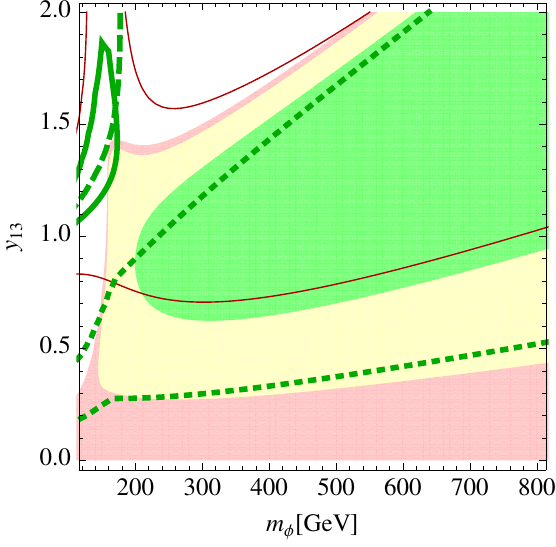}
  \caption{$t\bar t$ production constraints on the scalar isodoublet model in approach A (left plot) and B (right plot): FBA (inclusive in left plot and binned in right plot) at $1\sigma$ ($2\sigma$) in thick full green line (thin full red line), inclusive CA at $1\sigma$ ($2\sigma$) in thick dashed green line (thin dashed red line). In left plot $\sigma_{\rm TEV}$ at $1\sigma$ ($2\sigma$) in thick green dotted line (thin red dotted line), $\sigma_{\rm LHC}$ at $1\sigma$ ($2\sigma$) in thick green dash-dotted line (thin red dash-dotted line). In right plot CDF $m_{t\bar t}$ spectrum at $1\sigma$ ($2\sigma$) in thick green dotted line (thin red dotted line).  Parameter regions where the model can improve the SM $\chi^2$ by $-\Delta \chi^2 > 0,1,4$ are shaded in red, yellow and green respectively. Finally in left plot for $m_{\phi}<m_t$, the contours of constant branching fraction ${\rm Br}(t\to u \phi^0)$ are displayed in thin gray dashed lines.} 
  \label{fig:doublet_bounds}
\end{figure}

In this model the parameter scan is performed over the $\bar u_R t_L$ coupling ($y_{13}$) and mass ($m_\phi$) of the neutral isodoublet component $\phi^0$.\footnote{We do not consider the potential $d\bar d \to t\bar t$ contributions mediated by the charged isodoublet component, which are suppressed by smaller parton luminosities.} The resulting parameter fits for approach A and B are presented in Fig.~\ref{fig:doublet_bounds}.
We observe that in approach A (using $\sigma_{\rm TEV}$ and $\sigma^h_{\rm TEV}$) there is a small preferred parameter region at low $\phi^0$ masses ($m_\phi\lesssim m_t$, consistent with findings of~\cite{doublet}). 
Interestingly, $m_\phi$ is bounded from below by the LHC inclusive cross-section measurement. This is because the new $t\to \phi^0 u$ decay channel tends to suppress the reconstructed cross-sections (see Eq.~\eqref{eq:Br}) irrespective of the production mechanism inducing a tension between the Tevatron and LHC results.
When considering instead the complete CDF $m_{t\bar t}$ spectrum (approach B), the $m_{\phi}\lesssim m_t$ region is no longer favored. This happens because a light scalar exchanged in the $t-$channel tends to harden the $d\sigma/dm_{t\bar t}$ spectrum compared to SM QCD predictions, inducing a tension with the binned spectrum measurement by CDF (in this particular case, the discrepancy is most pronounced in the lowest $m_{t\bar t}$ bin which was measured precisely by CDF~\cite{CDFmtt}). Given our rough treatment of the correlated uncertainties entering $d\sigma_{\rm TEV}/dm_{t\bar t}$ and also the neglect of subleading SM contributions~\cite{Manohar:2012rs}, we shall keep the low mass region, disfavored by the $d\sigma/dm_{t\bar t}$ observables, in our analysis.
What remains in addition is a less interesting intermediate to heavy $\phi^0$ mass region ($m_\phi> 200$~GeV) where the FBA cannot be reproduced at the $1\sigma$ level however the overall agreement with the present experimental data is improved with respect to the SM (in order to clearly distinguish it from the light $\phi^0$ region, we shall only consider parameter values where the overall fit improvement with respect to the SM is at least $-\Delta \chi^2 >  4$, shaded green in right plot of Fig.~\ref{fig:doublet_bounds}).  
Finally in this model the tension between the inclusive FBA and CA measurements can be reduced below the $2\sigma$ level, even though the $68\%$ CL regions for both observables cannot be reached simultaneously.

\subsection{Scalar color triplet and sextet}

\begin{figure}[!t]
  \centering
  \includegraphics[width=0.47\textwidth]{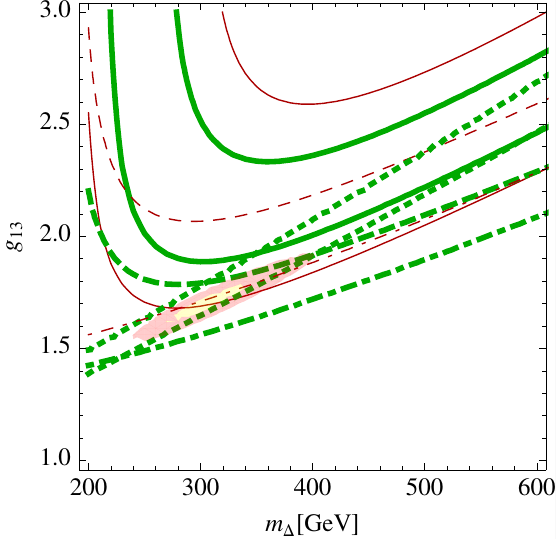}
  \includegraphics[width=0.47\textwidth]{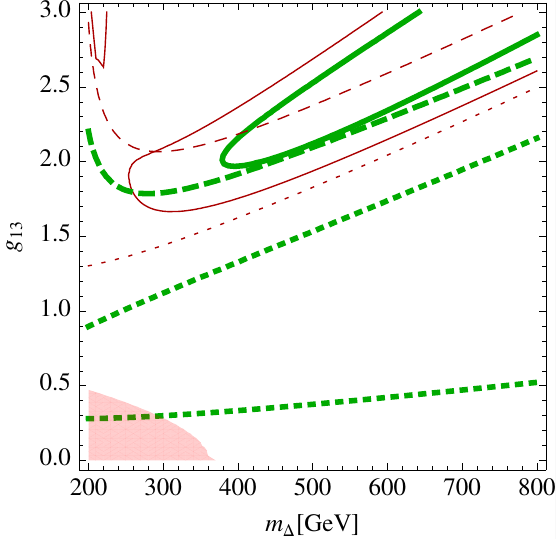}
  \caption{$t\bar t$ production constraints on the scalar color triplet model in approach A (left plot) and B (right plot): FBA (inclusive in left plot and binned in right plot) at $1\sigma$ ($2\sigma$) in thick full green line (thin full red line), inclusive CA at $1\sigma$ ($2\sigma$) in thick dashed green line (thin dashed red line). In left plot $\sigma_{\rm TEV}$ at $1\sigma$ ($2\sigma$) in thick green dotted line (thin red dotted line), $\sigma^h_{\rm TEV}$ at $2\sigma$ in thin red dash-dotted line. In right plot CDF $m_{t\bar t}$ spectrum at $1\sigma$ ($2\sigma$) in thick green dotted line (thin red dotted line).  Parameter regions where the model can improve the SM $\chi^2$ by $-\Delta \chi^2 > 0,1$ are shaded in red and yellow respectively.} 
  \label{fig:triplet_bounds}
\end{figure}

\begin{figure}[!t]
  \centering
  \includegraphics[width=0.47\textwidth]{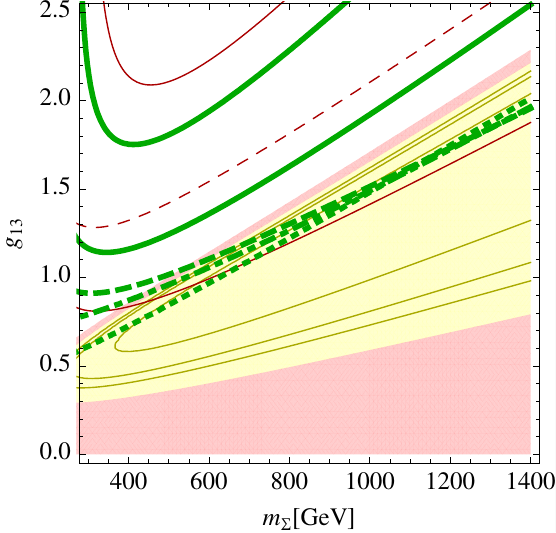}
  \includegraphics[width=0.47\textwidth]{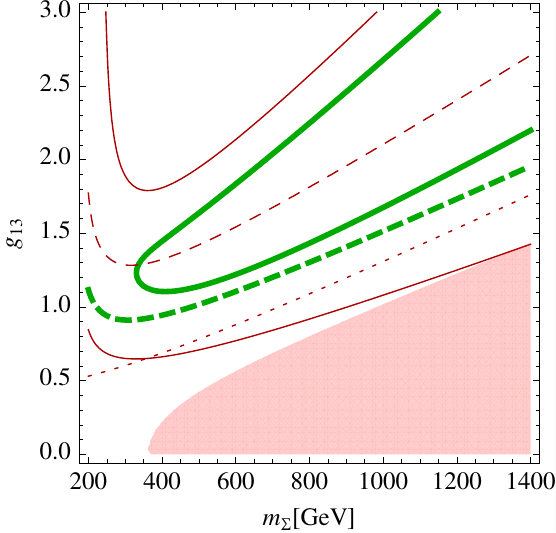}
  \caption{$t\bar t$ production constraints on the scalar color sextet model in approach A (left plot) and B (right plot): FBA (inclusive in left plot and binned in right plot) at $1\sigma$ ($2\sigma$) in thick full green line (thin full red line), inclusive CA at $2\sigma$ in thin dashed red line. In left plot $\sigma_{\rm TEV}$ at $1\sigma$  in thick green dotted line, $\sigma^h_{\rm TEV}$ at $1\sigma$ in thick green dash-dotted line. In right plot CDF $m_{t\bar t}$ spectrum at $2\sigma$ in thin red dotted line.  Parameter regions where the model can improve the SM $\chi^2$ by $-\chi^2 > 0, 1$ are shaded in red and yellow respectively.} 
  \label{fig:sextet_bounds}
\end{figure}

We consider chiral couplings of $\Delta$ and $\Sigma$ to right-handed di-quark currents $\bar u_R t_R^C$ with a coupling constant $g_{13}$. The results of the corresponding parameter fits are presented in Figs.~\ref{fig:triplet_bounds} and~\ref{fig:sextet_bounds}. 
Both of these models experience a tension between the FBA and the $t\bar t$ spectrum measurements~\cite{Gresham}, in particular at high $m_{t\bar t}$. While in approach A an overall improvement  in the goodness of fit can still be obtained in some parameter regions of the two models (see left plots in Figs.~\ref{fig:triplet_bounds} and~\ref{fig:sextet_bounds}); taking into account the complete binned $m_{t\bar t}$ spectrum measurement by CDF removes these preferred regions and the FBA discrepancy can no longer be addressed. Again, taking into account the possible caveats of the $d\sigma/dm_{t\bar t}$ constraints, we shall keep the $\Delta$ and $\Sigma$ model parameter regions preferred in approach A in our analysis of top spin observables. 

%
\section{TOP SPIN OBSERVABLES}
\label{sec:Spin}
%

\subsection{Definitions}
\label{sec:Spin:Def}

Since the spin information of a decaying top quark is not diluted by hadronization it is possible to investigate various top-spin polarization effects at hadron colliders. The starting point is the calculation of the production spin density matrices for the dominant partonic production channels of top quark pairs,
\begin{eqnarray}
&& q(p_1) + \overline{q}(p_2) \stackrel{}{\to} t(k_1,s_t) + \overline{t}(k_2,s_{\overline{t}}) \,,\nonumber \\
&& g(p_1) + g(p_2) \stackrel{}{\to} t(k_1,s_t) + \overline{t}(k_2,s_{\overline{t}}) \,,
\end{eqnarray}
via the relevant scattering amplitudes $\mathcal M_I$ where $I = q \overline{q}, gg $. 
The differential cross sections for these partonic processes can be written as $d \sigma_I (s_t, s_{\overline{t}} ) = \Phi_I |\mathcal M_I|^2 d\Gamma_{t\bar t}$ where $\Phi_I$ is the relevant initial state flux normalization and $d\Gamma_{t\bar t}$ is the $t\bar t$ phase space differential. The problem is thus reduced to evaluating the absolute squares of the relevant polarized scattering amplitudes
\begin{eqnarray}
|\mathcal M_I|^2 = \frac{1}{4} {\rm Tr} [ \rho_I ( {\bf 1}  + \mathbf{\hat{s}}_t  \cdot \boldsymbol{\sigma} )\otimes ( {\bf 1} + \mathbf{\hat{s}}_{\overline{t}} 
\cdot \boldsymbol{\sigma}) ]\,,
\end{eqnarray}
where $\rho_I$ is the corresponding partonic production spin density matrix describing the production of (on-shell) top quark pairs in a specific spin configuration, while $\mathbf{\hat{s}}_t$ ($\mathbf{\hat{s}}_{\overline{t}}$) is the unit polarization vector of the top (anti top) quark in its rest frame and $\boldsymbol\sigma = (\sigma_1,\sigma_2,\sigma_3)^T$ is a vector of Pauli matrices.

Conveniently, one starts with the most general decomposition of the spin density matrix $\rho$ (suppressing the initial state subscript $I$) for the production of a top quark pair\:
\begin{eqnarray}
\rho = A\, {\bf 1} \otimes  {\bf 1} + B_i^{t} \, \sigma_i  \otimes {\bf 1} +  B_i^{\overline{t}} \, {\bf 1} \otimes \sigma_i + C_{ij} \, \sigma_i \otimes \sigma_j \, , 
\label{eq:rho}
\end{eqnarray}
where the functions $A$, ${B}_i^t$ (${B}_i^{\overline{t}}$) and $C_{ij}$ describe the spin-averaged production cross section, polarization of
a top (anti top) quarks and the top - anti top spin-spin correlations, respectively.
Using the spin four-vectors defined as
\begin{eqnarray}
s_t^{\mu} &=&  \left ( \frac{ \mathbf{k}_1 \cdot \mathbf {\hat{s}}_t}{m_t}, \mathbf {\hat{s}}_t + \frac{ \mathbf{k}_1 (\mathbf{k}_1 \cdot \hat{\mathbf s}_t)}{ m_t (E_t + m_t)} \right )\,,
\nonumber \\
s_{\overline{t}}^{\mu} &=&  \left (  \frac{ \mathbf{k}_2\cdot \mathbf{\hat{s}}_{\overline{t}}}{m_t}, \mathbf{\hat{s}}_{\overline{t}} + \frac{ \mathbf{k}_2 (\mathbf{k}_2 \cdot \mathbf{\hat{s}}_{\overline{t}})}{ m_t (E_{\bar t} + m_t)} \right )\,,
\end{eqnarray}
the decomposition of the squared scattering amplitude $|\mathcal M|^2$ can be written as
\begin{eqnarray}
|\mathcal M|^2 = a + b^{t}_\mu s_t^{\mu} + b^{\overline{t}}_\mu s_{\overline{t}}^{\mu} + c_{\mu\nu} s_t^{\mu} s_{\overline{t}}^{\nu}\,,
\label{eq:M2}
\end{eqnarray}
and by comparing expressions \eqref{eq:rho} and \eqref{eq:M2} one can extract the functions $A$, ${B}_i^t$ (${B}_i^{\overline{t}}$) and $C_{ij}$.
With this at hand, the various top spin observables $\langle {\cal O} \rangle$ can be calculated as
\begin{eqnarray}
 \langle {\cal O} (\mathbf{{S}}_t, \mathbf{{S}}_{\overline{t}}) \rangle_I = \frac{\Phi_I}{\sigma_I} \int d\Gamma_{t\bar t} { {\rm Tr} [ \rho_I \cdot {\cal O} (\mathbf{{S}}_t, \mathbf{{S}}_{\overline{t}} ) ]} 
\,,
\end{eqnarray}
where $\sigma_I = \Phi_I \int d\Gamma_{t\bar t} {\rm Tr}[\rho_I]$ is the unpolarized production cross-section and  $\mathbf{{S}}_t = \boldsymbol{\sigma}/2 \otimes {\bf 1} \, ( \mathbf{{S}}_{\overline{t}} = {\bf 1} \otimes \boldsymbol{\sigma}/2)$ is the top (anti top) spin operator.  
In particular, we consider the following spin observables
\begin{eqnarray}
{\cal O}_1 &=&  \mathbf{{S}}_t \cdot \mathbf{S}_{\overline{t}} \,,
\nonumber \\
{\cal O}_2 &=&  \mathbf{{S}}_t \cdot \mathbf{\hat{a}}\,, \quad \bar{\cal O}_2 = \mathbf{{S}}_{\bar t} \cdot \mathbf{\hat{b}} \,,
\nonumber \\
{\cal O}_3 &=& 4 ( \mathbf{{S}}_t \cdot \mathbf{\hat{a}} ) ( \mathbf{S}_{\overline{t}} \cdot \mathbf{\hat{b}} )\,,
\end{eqnarray}
which give the net spin polarization of the top - anti top system, polarization of the (anti) top quark, and the top - anti top spin correlation both with respect to spin quantization axes $\mathbf{\hat a}$ and $\mathbf{\hat b}$, respectively.
At the parton level ${\cal O}_3$ is related to the spin correlation function $C_{ij}$ in Eq.~\eqref{eq:rho}, namely
\begin{eqnarray}
\braket{ {\cal O}_3 } = \frac{ \sigma_{t \bar{t}} (\uparrow \uparrow) + \sigma_{t \bar{t}} (\downarrow \downarrow) - \sigma_{t \bar{t}} (\uparrow \downarrow)- 
\sigma_{t \bar{t}} (\downarrow \uparrow)}
{\sigma_{t \bar{t}} (\uparrow \uparrow) + \sigma_{t \bar{t}} (\downarrow \downarrow) + \sigma_{t \bar{t}} (\uparrow \downarrow)+ \sigma_{t \bar{t}} (\downarrow \uparrow)}\,,
\end{eqnarray}
where the arrows refer to the up and down spin orientations of the top and the anti top quark with  respect to the $\mathbf{\hat a}$ and $\mathbf{\hat b}$ quantization axes, respectively.
It can be measured using the double differential angular distribution of the top and anti top quark decay products:
\begin{eqnarray}
\frac{1}{\sigma} \frac{d^2 \sigma}{d \cos\theta_f d\cos\theta_{\bar f}} = \frac{1}{4} \left ( 1 + B_t \cos\theta_f + B_{\bar{t}} \cos\theta_{\bar f} - C \cos\theta_f \cos\theta_{\bar f} \right )\,,
\label{eq:diffsigma}
\end{eqnarray}
where $\theta_f (\theta_{\bar f})$ is the angle between the direction of the top (anti top) spin analyzer $f,\,(\bar f)$ (which can be either a direct $t$ ($\bar t$) daughter $W^+,b$ ($W^-, \bar{b}$) or a $W^+ (W^-)$ decay product
$\ell^+ (\ell^-), \nu (\bar \nu)$ or jets) in the $t$ $(\bar{t})$ rest frame and the $\hat{\bf a}$ ($\hat{\bf  b}$) direction in the $t \bar{t}$ center of mass frame (when the corresponding frame transformation is a rotation free boost, c.f.~\cite{spinSM}). Analogously ${\cal O}_2$ and $\bar{\cal O}_2$ are related to the (anti)top spin polarization coefficients $B_t$ and $B_{\bar t}$. We note in passing that in absence of CP violation $B \equiv B_t = \mp B_{\bar t}$ for $\hat{\bf a} = \pm \hat{\bf b}$. For perfect (anti)top spin analyzers whose flight directions are $100\%$ correlated with the directions of the (anti)top spin then
\beq
\braket{\mathcal O_3} = C\,, \quad \braket{\mathcal O_2} = B_t\,,\quad \braket{\bar {\mathcal O}_2} = B_{\bar t}\,.
\eeq
This limit is a good approximation for the charged leptons from $W$ decays~\cite{Brandenburg:2002xr}. For other (anti)top spin analyzers one needs to apply the corresponding top spin analyzing power factors $\kappa_{f(\bar f)}$ (where $\kappa_{\ell^+(\ell^-)}\simeq1$) in Eq.~\eqref{eq:diffsigma} as
\begin{equation}
C \to C \kappa_{f} \kappa_{\bar f}\,, \quad B_{t(\bar t)} \to B_{t(\bar t)} \kappa_{f(\bar f)}\,.
\end{equation}
The values of $\kappa_{f(\bar f)}$ are presently known at NLO in QCD and can be found in~\cite{Brandenburg:2002xr}. 
 Finally,  ${\cal O}_1$ can be probed using the spin analyzer opening angle distribution
\begin{eqnarray}
\frac{1}{\sigma} \frac{d \sigma}{d \cos \phi} = \frac{1}{2} \left (  1 - D  \cos \phi \right )\,,
\end{eqnarray}
where $\phi$ is the angle between the direction of flight of the two (top and anti top) spin analyzers, defined in the $t$ and $\bar{t}$ frames, respectively. Again, for $f=\ell^+$, $\bar f = \ell^-$ one then obtains
\beq
\braket{\mathcal O_1} = D\,,
\eeq
while for other spin analyzers the appropriate $\kappa_{f(\bar f)}$ corrections should be applied. 

The arbitrary unit vectors $\bf{\hat a}$ and $\bf{\hat b}$ specify different spin quantization axes which can be chosen to maximize the desired polarization and correlation effects.
We work with the following choices:
\begin{align}
\hat{\bf a} &= - \hat{\bf b} = \hat{\bf k}_1\,,  && ({\rm ``helicity" \; basis})\,,&&
\nonumber \\
\hat{\bf a} &= \hat{\bf b} = \hat{\bf p}\,,  && ({\rm ``beamline" \; basis})\,,&&
\nonumber \\
\hat{\bf a} &= \hat{\bf b} = \hat{\bf d}_X\,,  && ({\rm ``off-diagonal" \; basis,\; specific\; for\;  model\;  X})\,,&&
\label{eq:axes}
\end{align}
where $\hat{\bf p}$ is the direction of the incoming beam and $\hat{\bf k}_1$ is the direction of the outgoing top quark, both in the $t\bar t$ center of mass frame.

By the detailed study of the top (anti top) decay products one can obtain valuable information about top spin observables and use them to distinguish among the different models addressing the FBA puzzle. In order to maximize top spin effects it is advisable to choose a proper spin quantization axis. 
For the leading order QCD quark-antiquark annihilation dominating the $t\bar t$ production at the Tevatron, a special off-diagonal axis was shown to exist~\cite{Parke:1996pr}, for which the top spins are $100\%$ correlated. It is given by quantizing the spins with the axis $ \hat {\bf d}_{q\bar q,\rm SM}$ determined as
\begin{eqnarray}
\hat{\bf d}_{q\bar q, \rm SM} = \frac{ - \hat{\bf p}  + ( 1 - \gamma) z \; \hat{\bf k}_1}{\sqrt{ 1 - (1 - \gamma^2) z^2}}\,,
\label{eq:dSM}
\end{eqnarray}
where $z = \hat{\bf p} \cdot \hat{\bf k}_1 = \cos \theta$ and $\gamma = E_t/m_t = 1/\sqrt{1 - \beta^2}$ and interpolates between the beamline basis at the threshold ($\gamma \to 1$) and the helicity basis for  ultrarelativistic energies ($\gamma \to \infty$). 
There is no such optimal axis for the gluon-gluon fusion process (dominating the $t\bar t$ production at the LHC)\footnote{At low $m_{t\bar t}$ the top quark pair production via 
gluon-gluon fusion is dominated by like-helicity gluons. Consequently, spin correlations are maximal in the helicity basis~\cite{Mahlon:2010gw}.}, but
it is always possible to find a basis, in which spin correlations are maximal.
A general procedure for finding such an off-diagonal basis is given in~\cite{Mahlon:1997uc, Uwer:2004vp}.
The idea is to determine the maximal eigenvalue of the matrix function $C_{ij}$ in Eq.~\eqref{eq:rho} and the corresponding eigenvector, which provides the off-diagonal quantization axis $\hat {\bf d}_X$.
Such an off-diagonal axis can be constructed for any model. As an example we present the explicit formula for the axigluon model (see also~\cite{Bai:2011uk})
\begin{eqnarray}
\hat{\bf d}_{ A} = \frac{ - \hat{\bf p}  + \left ( ( 1 - \gamma) z  -  g_A^2 \sqrt{\gamma^2-1} \frac{s}{s - m_A^2} \right )\, \hat{\bf k}_1}
{\sqrt{ 1 - (1 - \gamma^2)z^2 + 2 \gamma \sqrt{\gamma^2-1} g_A^2 s/(s - m_A^2) z - (1 - \gamma^2) g_A^4 s^2/(s - m_A^2)^2}}\,,
\label{eq:axigax}
\end{eqnarray}
where $g_A$ and $m_G$ are free parameters of the axis. The $q\bar q$ annihilation induced top spin correlations in this basis are maximal when these parameters match the actual axigluon resonance parameters (i.e. when $g_A = g_A^u g_A^t$ and $m_A = m_G$). It turns out however, that such custom axes are not necessarily better in discriminating between the SM and NP contributions to top pair production via $q\bar q$ annihilation. This can be easily understood by treating the SM and SM+NP as two competing hypotheses, among which we wish do discriminate using top spin correlations. Then the SM $q\bar q$ off-diagonal axis  is already optimal for one of the hypotheses (it maximizes the leading QCD $q\bar q$ induced spin correlations) and is furthermore completely fixed by known SM parameters. Conversely any NP model off-diagonal axis needs to reduce to the SM one in the  limit where the NP parameters determining the NP off-diagonal basis decouple ($g_A\to 0$ and/or $m_A \to \infty$ for the axigluon axis in Eq.~\eqref{eq:axigax}). In the case of the axigluon model off-diagonal axis, we have checked explicitly, that in the interesting regions of parameter space and for arbitrary values of $g_A,m_A$, the axigluon predictions do not deviate from SM values significantly more than in the helicity or the SM $q\bar q$ off-diagonal axis. Consequently, in our numerical study we use the helicity (for spin correlation $C_{\rm hel}$ and polarization $B_{\rm hel}$ coefficients), beamline (for $C_{\rm beam}$, $B_{\rm beam}$) and SM $q\bar q$ off-diagonal (for $C_{\rm off}$, $B_{\rm off}$) spin quantization axes defined in Eqs.~\eqref{eq:axes} and~\eqref{eq:dSM} respectively. 

\subsection{Results}

\begin{figure}[!t]
  \centering
  \includegraphics[width=0.4\textwidth]{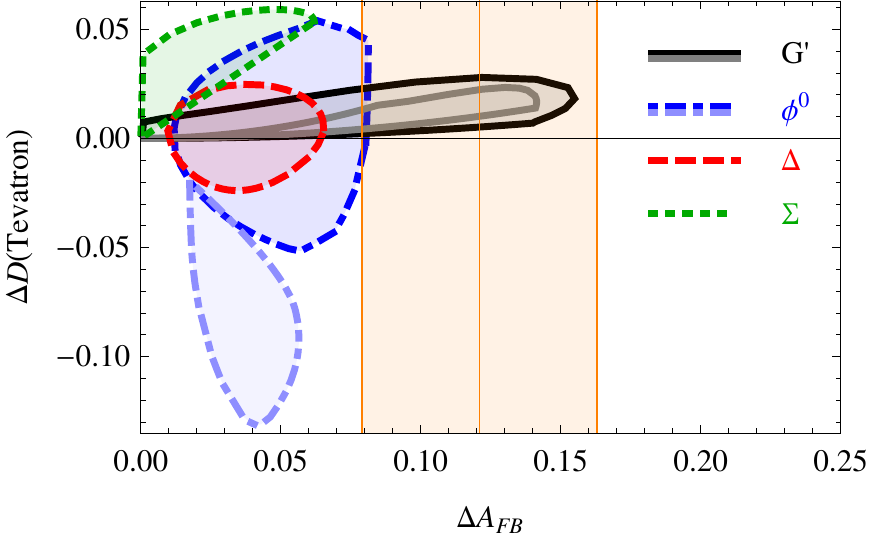}
  \includegraphics[width=0.4\textwidth]{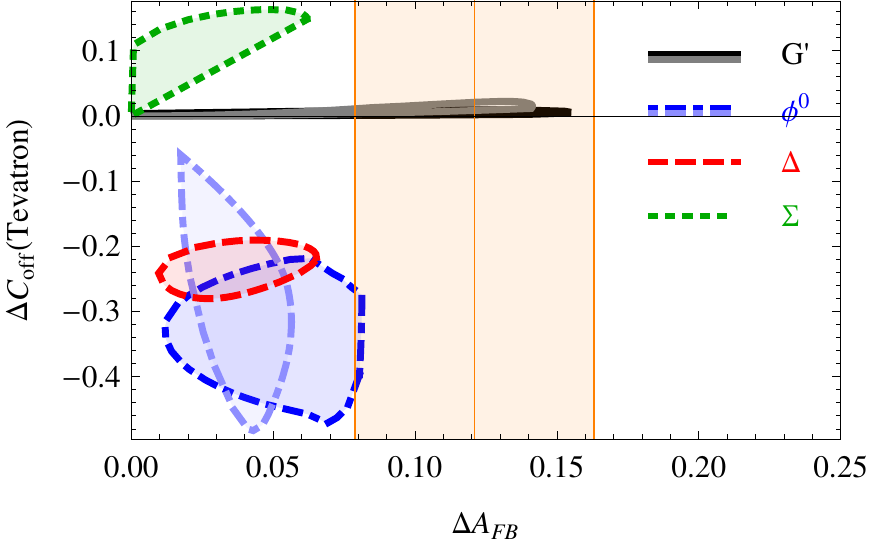}
\\
  \includegraphics[width=0.4\textwidth]{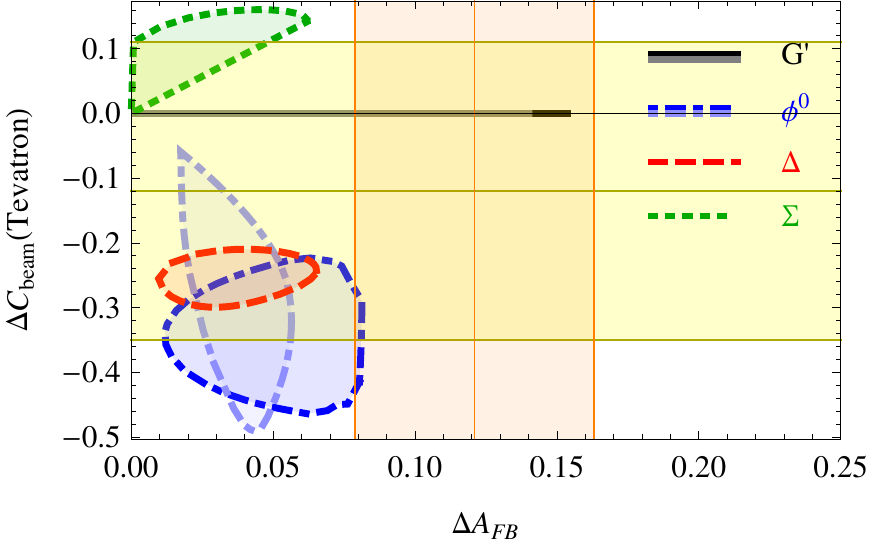}
  \includegraphics[width=0.4\textwidth]{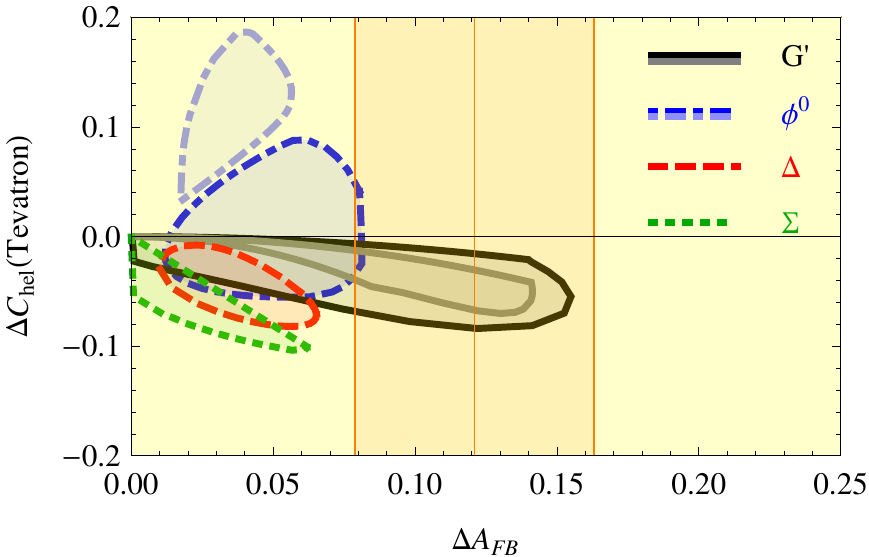}
  \\
  \includegraphics[width=0.4\textwidth]{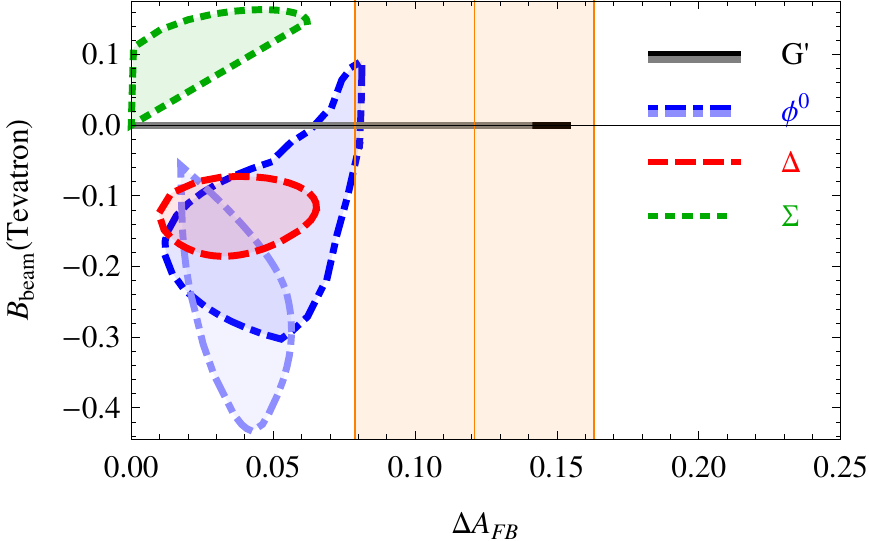}
  \includegraphics[width=0.4\textwidth]{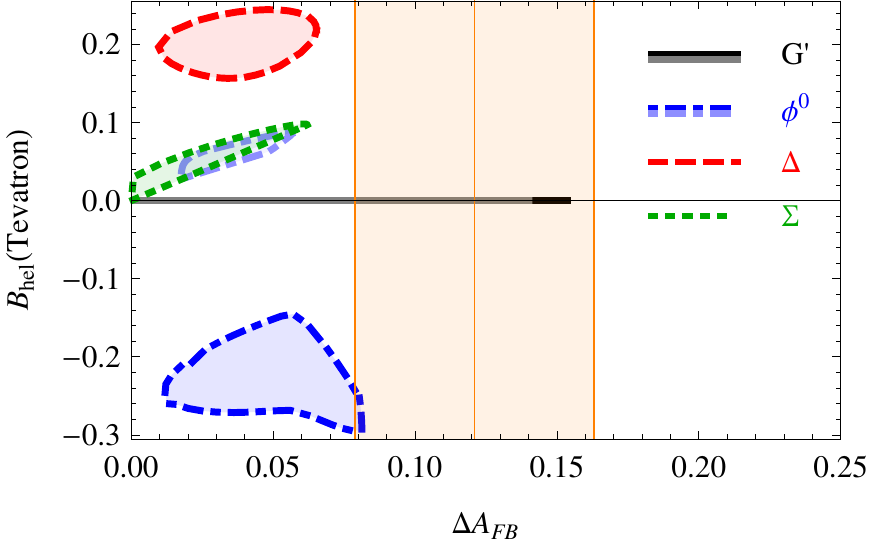}
  \caption{Correlations between the NP contributions to the inclusive FBA and various spin observables at the Tevatron (see text for details and definitions). The present experimental results ($68\%$ C.L. regions) are shaded in horizontal and vertical bands. The NP model predictions are determined from the global fit as specified in Sec.~\ref{sec:fit} and are bounded by full (axigluon $G'$ in the low ($m_G\lesssim 450$~GeV in black) and high ($m_G\gtrsim 700$~GeV in gray) mass regions), dashed (scalar color triplet $\Delta$), dotted (scalar color sextet $\Sigma$) and dot-dashed (neutral component of the scalar isodoublet $\phi^0$ in the low ($m_\phi\lesssim m_t$ in darker shade) and high ($m_\phi > 200$~GeV in lighter shade) mass region) contours. } 
  \label{fig:spinTev}
\end{figure}

\begin{figure}[!t]
  \centering
  \includegraphics[width=0.4\textwidth]{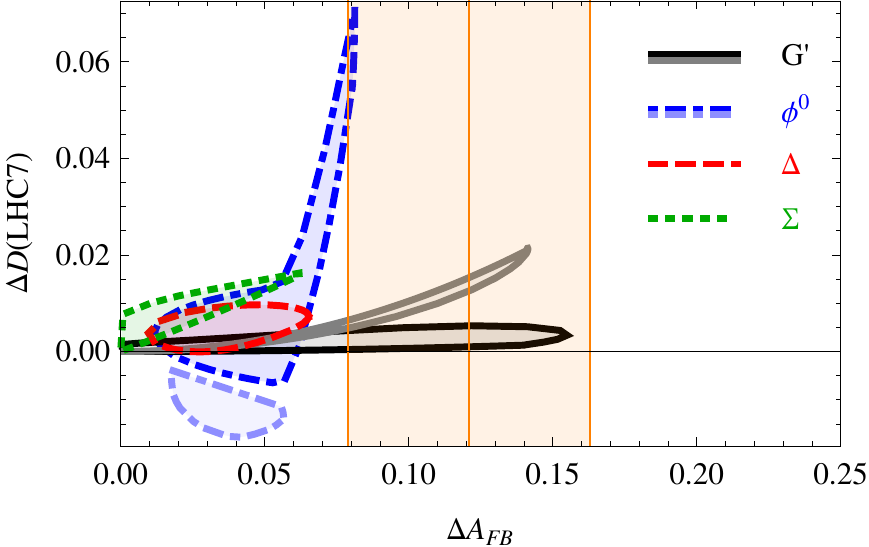}
  \includegraphics[width=0.4\textwidth]{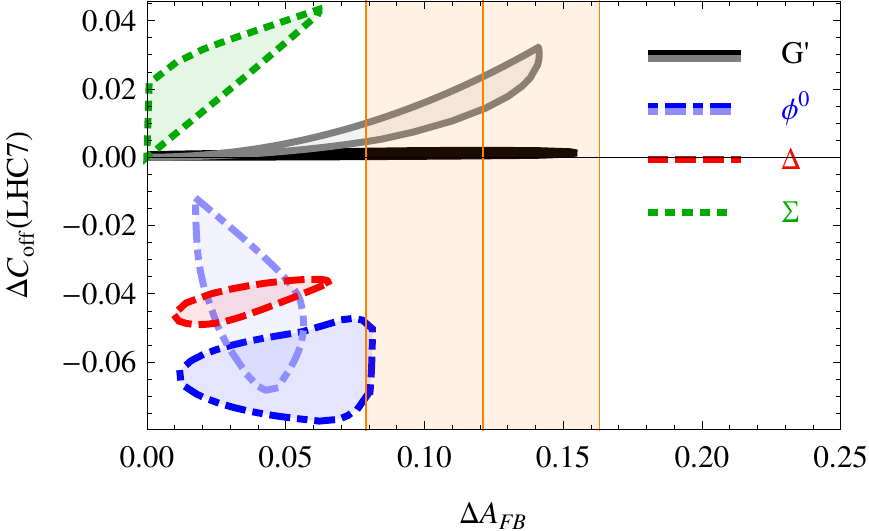}
\\
  \includegraphics[width=0.4\textwidth]{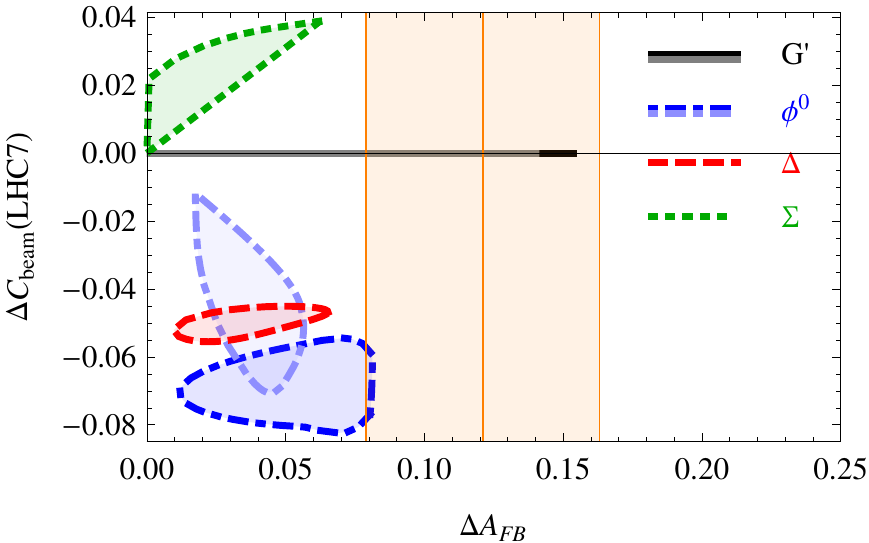}
  \includegraphics[width=0.4\textwidth]{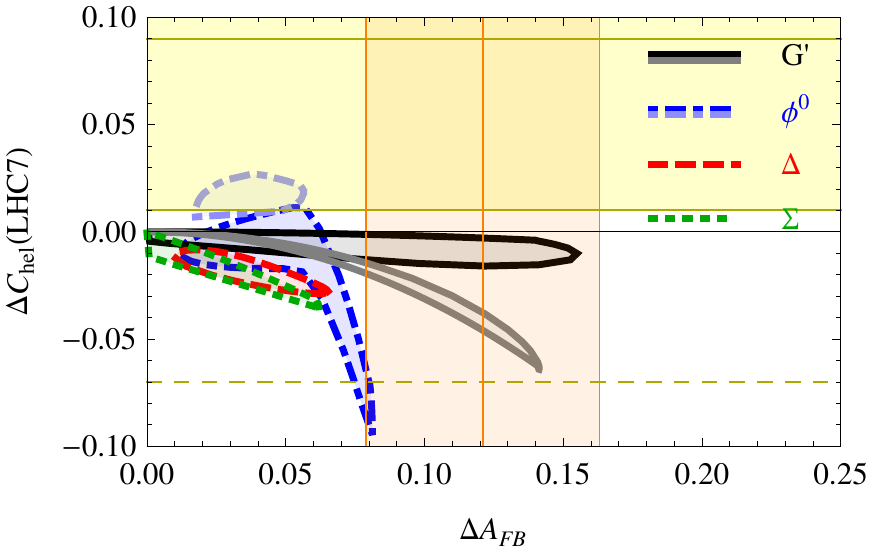}
  \\
  \includegraphics[width=0.4\textwidth]{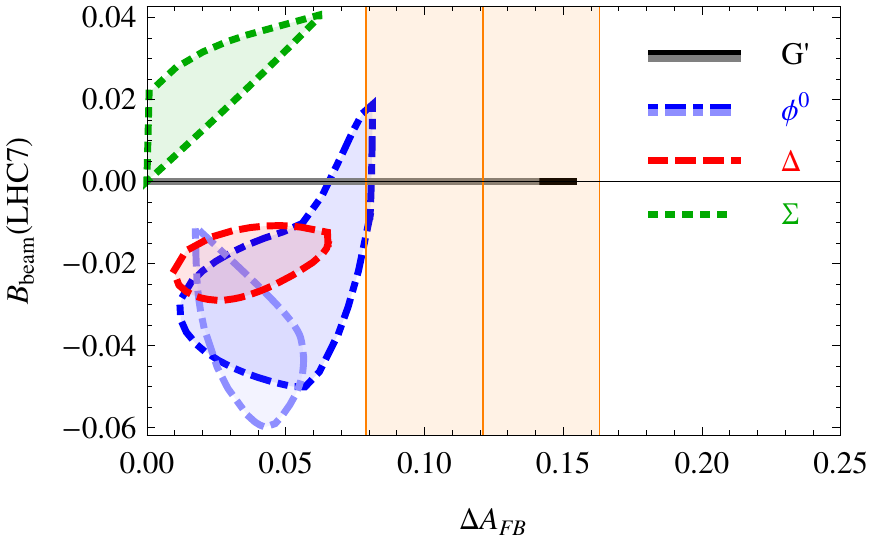}
  \includegraphics[width=0.4\textwidth]{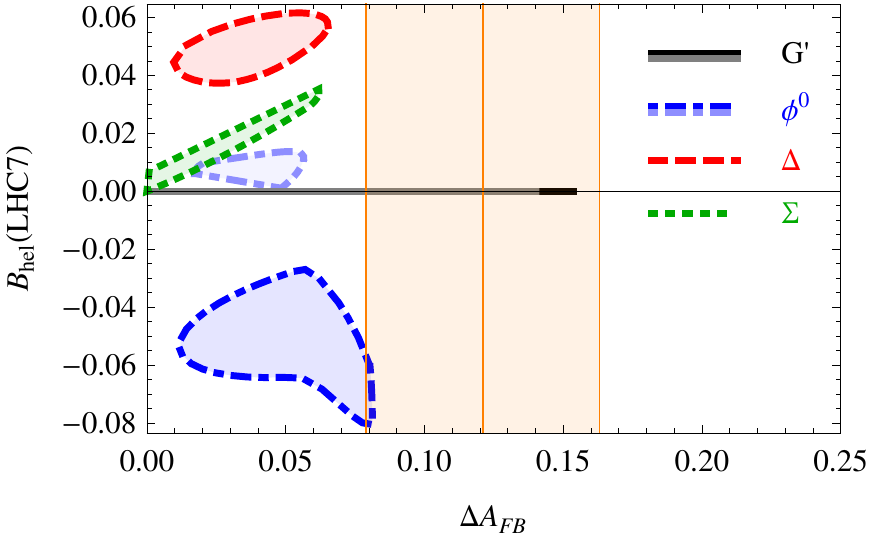}
  \caption{Correlations between the NP contributions to the inclusive FBA at the Tevatron and various spin observables at the 7~TeV LHC (see text for details and definitions). The present experimental results ($68\%$ C.L. regions) are shaded in horizontal and vertical bands. For $\Delta C_{\rm hel}$ we also show the $95\%$ C.L. contour in thin dashed line. The NP model predictions are determined from the global fit as specified in Sec.~\ref{sec:fit} and are bounded by full (axigluon $G'$ in the low ($m_G\lesssim 450$~GeV in black) and high ($m_G\gtrsim 700$~GeV in gray) mass regions), dashed (scalar color triplet $\Delta$), dotted (scalar color sextet $\Sigma$) and dot-dashed (neutral component of the scalar isodoublet $\phi^0$ in the low ($m_\phi\lesssim m_t$ in darker shade) and high ($m_\phi > 200$~GeV in lighter shade) mass region)  contours.} 
  \label{fig:spinLHC7}
\end{figure}

In this section we present predictions for the various  top spin observables at the Tevatron as well as the $7$~TeV (and $8$~TeV) LHC within the various NP model parameter regions which are able to address the FBA puzzle, as determined in Sec.~\ref{sec:fit}. In particular we present correlations between the inclusive and high $m_{t\bar t}$ FBA values as measured at the Tevatron, and the shifts of the various spin observables from their corresponding SM values. We define (see Sec.~\ref{sec:obs}) $\Delta A_{FB} \equiv A_{FB} - A_{FB}^{\rm SM}$, $\Delta C_i \equiv C_i - C_i^{\rm SM}$ and $\Delta D \equiv D - D^{\rm SM}$. On the other hand since QCD produced top quarks are not polarized, (neglecting tiny electroweak contributions) we assume $B_i^{\rm SM}\simeq 0$ and present results for $B_i$ in presence of NP directly.  
The predictions for the relevant spin observables at the Tevatron are shown in Fig.~\ref{fig:spinTev}.
First note that the results for the SM $q\bar q$ off-diagonal axis at the Tevatron turn out to be almost identical to the beamline axis (and very similar at the LHC, see Fig.~\ref{fig:spinLHC7}). Both bases provide good potential discrimination between color sextet on one hand, and color triplet or isodoublet scalar models on the other hand. The off-diagonal basis exhibits marginally better sensitivity only for the axigluon ($G'$) model. However, since purely axial couplings of $G'$ to quarks do not produce polarized top quarks, $B_i$ vanishes for the axigluon model and consequently we do not plot $ B_{\rm off}$ dependence separately.

We observe that existing spin observable measurements at the Tevatron do not overly constrain selected NP models. Some sensitivity to the light scalar isodoublet model is exhibited by the recent beamline axis spin correlation measurement by D\O~\cite{spinTEV1} as seen in the center left plot in Fig.~\ref{fig:spinTev}. 
On the other hand (anti)top polarization ($B_i$ both in the beamline and in the helicity basis) offers a very powerful probe of scalar $t$-channel models and a $\mathcal O(20\%)$ precision measurement (in helicity basis) could already test (and discriminate between) the scalar color triplet ($\Delta$) and isodoublet ($\phi^0$) model explanations of the FBA.  
Finally, the axigluon ($G'$) models in general give very small contributions to the chosen spin observables. For example, at the Tevatron, spin correlation measurements at  $\mathcal O(2\%)$ precision  would be required to probe such FBA explanations.

The results for the relevant spin observables at the 7~TeV LHC are shown in Fig.~\ref{fig:spinLHC7}.\footnote{The results for $\Delta D$, $\Delta C_i$ and $B_i$ at the 7~TeV and 8~TeV LHC are almost identical and we do not show the later separately.}
Among these, presently the most powerful probe of FBA inspired models is the helicity basis spin correlation as measured recently by ATLAS~\cite{Aad:2012sm}. In particular it already represents a non-trivial constraint for the scalar isodoublet and heavy axigluon models. In the light scalar isodoublet scenario, the large negative deviation in $\Delta C_{\rm hel}$ can be traced to sizable non-standard $t\to \phi^0 u$ decay rate, reducing the experimental $t\bar t$ reconstruction efficiency (which is then compensated by enhancing the $ u \bar u \to  t \bar t$ cross-section contribution). In term this leads effectively to a reduced gluon fusion component in the $t\bar t$ cross-section which in this basis contributes sizable positive spin correlations in the SM.

Comparably sensitive (and presently unmeasured) observables are also the opening angle spin correlation coefficient $D$, beamline axis spin correlations ($C_{\rm beam}$) and helicity axis top polarization ($B_{\rm hel}$). $\mathcal O(5\%)$ precision measurements of these quantities could discern among the $t$-channel scalar models. Finally, again we observe that the axigluon ($G'$) model (especially in the light $m_G\lesssim 450$~GeV region) gives very small contributions to the chosen spin observables at the LHC and will be difficult to probe in this way. 

%
\section{Conclusions}
\label{sec: conclusions}
%

We have performed a comprehensive analysis of $t\bar t$ production phenomenology at the Tevatron and the LHC within effective single NP resonance models addressing the FBA puzzle. 
We have quantified an inherent tension between the large positive FBA measurements  at the Tevatron and precise CA measurements at the LHC (consistent with zero). In particular the later conclusively exclude the $W'$ and $Z'$ explanations of the FBA anomaly. Among the considered models, only a color octet axial-vector axigluon state (of mass $m_G \sim 400$~GeV or $m_G\gtrsim 1$~TeV) can reproduce the central experimentally determined values of the inclusive and the high $m_{t\bar t}$ bin FBA without being in severe conflict with other $t\bar t$ constraints. In addition, a light scalar isodoublet ($m_\phi \lesssim m_t$) model predictions can barely reach the one sigma region for the inclusive FBA average, while the central value is in conflict with the LHC cross-section and CA measurements and also the recent top spin correlation measurement by ATLAS.  Finally, the  scalar color triplet and sextet models are constrained by cross-section and $m_{t\bar t}$ spectrum measurements at the Tevatron. However, given the caveats associated with properly evaluating the contributions of these models (and also the isodoublet model) to the existing experimentally reconstructed $m_{t\bar t}$ spectra (unfolded using SM signal templates), we suggest the experiments at the Tevatron and the LHC perform dedicated studies to settle the issue. 

For the favored NP model parameter regions we have derived predictions of (anti)top spin polarization and $t\bar t$ spin correlation observables at the Tevatron and the LHC. All scalar models addressing the FBA puzzle predict significant deviations in several top spin observables. At the Tevatron the most promising are the (anti)top polarization fractions with respect to the helicity spin quantization axis, which can deviate by more than $20\%$ from their SM predicted values and their measurement could discriminate between scalar isodoublet and color triplet models. At the $7$ (and $8$)~TeV LHC the scalar $t-(u-)$channel models predict smaller effects, but helicity axis top spin correlation and polarization measurements at the $5-10\%$ accuracy can yield competitive constraints, as exemplified by the recent ATLAS~\cite{Aad:2012sm} result. On the other hand, a light ($m_G\lesssim 450$~GeV) axigluon model predicts very small effects in top spin observables both at the Tevatron and at the LHC, and will be difficult to constrain in this way (much larger effects may be expected only if in addition to the axial component, a vector coupling to $q\bar q$ pairs is present~\cite{Falkowski:2011zr,AFBvsAC}).

While in the present top spin observables' analysis we have focused on inclusive $t\bar t$ samples, binned distributions of these quantities in $m_{t\bar t}$ and (anti)top rapidities could further enhance NP effects relative to the SM dominant $gg\to t\bar t$ subprocess and help to discriminate between various possible contributions (c.f.~\cite{Kagan:2011yx, Arguin:2011xm, Krohn:2011tw, Bai:2011uk, Falkowski:2011zr, Berger:2012nw} for recent proposals along these lines). 

\acknowledgments
J.F.K gratefully acknowledges useful discussions with Gilad Perez and Jure Zupan.
The work of S.F. and J.F.K. was supported in part  by the Slovenian Research Agency.

\appendix

%
\section{Partonic Amplitudes}
%

\subsection{{ Parametrization of Standard Model and New Physics Contributions}}

At the partonic level in the SM  the $q(p_1)  \bar q (p_2) \to  t(k_1,s_t) \bar t (k_2,s_{\bar t})$ amplitude is given by
\begin{equation}
i \mathcal M_{SM}^{q \bar{q}}(s_t,s_{\bar t}) = i \frac{g^2_s}{s}  \bar v(p_2) \gamma^\mu T^a u(p_1) \bar u( k_1,s_t) \gamma_\mu T^a v (k_2, s_{\bar t})\,,
 \label{ASMqq}
\end{equation}
with $s = (p_1 + p_2)^2 = (k_1 +k_2)^2$,  $t= (p_1- k_1)^2$ and $u = (p_1- k_2)^2$. Also, $g_s$ denotes the QCD coupling while $T^a$ are the Gell-Mann $SU(3)$ generator matrices. 

Then the unpolarized $q \bar{q}$ SM cross section is 
\begin{eqnarray}
\frac{d \sigma_{SM}^{q\bar{q}, \rm unpol.}}{d t}  = \frac{2 \pi \alpha_s^2}{9 s^2} [2 - (1 - \cos^2\theta) \beta^2 ]\,,
\end{eqnarray}
while for $gg$ initial state
\begin{eqnarray}
\frac{d \sigma_{SM}^{gg, \rm unpol.} }{d t}  = \frac{\pi \alpha_s^2}{48 s^2} \frac{[9 (1 + \beta^2 \cos^2\theta )-2]}{(1 - \beta^2 \cos^2\theta)^2} [ 1 + 2 \beta^2 (1-\beta^2) (1 - \cos^2\theta) - 
\beta^4 \cos^4\theta]\,,
\end{eqnarray}
where, as usual, $\beta = \sqrt{1 - 4 m_t^2/s}$ and $\theta$ is the scattering angle in the $t\bar{t}$ center of mass system. 

Since we are also interested in top spin observables, we present also the results for the polarized cross sections. 
The new physics models which we consider have particles which couple exclusively to quarks and antiquarks and therefore their amplitudes interfere only 
with the SM $q \bar{q}$ parts.  The cross sections receive the contribution from the SM, the interference between the SM and the NP, and the NP part only. 
For all models, denoted by $X$, we thus write
\begin{eqnarray}
\frac{d \sigma_{SM+ X}^{q\bar{q}, \rm pol.}}{d t}  = \frac{1}{16 \pi s^2} \frac{1}{4 N^2} ( |\mathcal M^{q\bar{q}, \rm pol.}_{SM}|^2 + |\mathcal M^{ \rm pol.}_{SM*X}|^2 + |\mathcal M^{ \rm pol.}_X|^2  ) \,,
\end{eqnarray}
where $N=3$ is the number of colors each square of the polarized amplitudes can be written as
\begin{eqnarray}
|\mathcal M^{ \rm pol.}|^2 &=& a + b^t_{\mu} s_t^{\mu}  + b^{\bar{t}}_{\mu} s_{\bar{t}}^{\mu} + c_{\mu\nu} s_t^{\mu} s_{\bar{t}}^{\nu} \,,
\end{eqnarray}
and further
\begin{eqnarray}
b^t_{\mu} s_t^{\mu} &=& b^t_1 (p_1 \cdot s_t) + b^t_2 (p_2 \cdot s_t)\,,
\\
b^{\bar t}_{\mu} s_{\bar t}^{\mu} &=& b^{\bar t}_1 (p_1 \cdot s_{\bar t}) + b^{\bar t}_2 (p_2 \cdot s_{\bar t})\,,
\\
c_{\mu\nu} s_t^{\mu} s_{\bar{t}}^{\nu} &=& c (s_t \cdot s_{\bar t}) + d [ (p_1 \cdot s_t)(p_1 \cdot s_{\bar t}) + (p_2 \cdot s_t)(p_2 \cdot s_{\bar t}) ] \,,
\nonumber \\
&& + e (p_1 \cdot s_t)(p_2 \cdot s_{\bar t}) + f (p_2 \cdot s_t)(p_1 \cdot s_{\bar t})\,.
\label{eq:Mpol}
\end{eqnarray}
Due to the CP invariance in our models, $b_1^{\bar t} = - b_2^t$ and $b_2^{\bar t} = - b_1^t$.

The spin coefficients for the polarized $q \bar q$ squared amplitude in the SM are as follows:
\begin{eqnarray}
a_{SM}^{q\bar q} &=& C_{SM}^{q\bar q} \left [ 2 - \beta^2 (1 - z^2) \right ]\,,
\nonumber \\
c_{SM}^{q\bar q} &=& C_{SM}^{q\bar q} \left [ \beta^2 (1 - z^2) \right ]\,,
\nonumber \\
e_{SM}^{q\bar q} &=& C_{SM}^{q\bar q} \left [- \frac{4}{s} (1 + \beta z) \right ]\,,
\nonumber \\
f_{SM}^{q\bar q} &=& C_{SM}^{q\bar q} \left [ - \frac{4}{s} (1 - \beta z) \right ]\,,
\end{eqnarray}
where $C_{SM}^{q\bar q} = 16 \pi^2 \alpha_s^2 (N^2 -1)$  and $z = \cos\theta$. All other coefficients are zero. 
On the other hand, the polarized SM $gg$-cross section has the form 
\begin{eqnarray}
\frac{d \sigma_{SM}^{gg, \rm pol.}}{d t}  = \frac{1}{16 \pi s^2} \frac{1}{4 (N^2-1)^2}  |\mathcal M^{gg, \rm pol.}_{SM}|^2\,,
\end{eqnarray}
and $|M^{gg, \rm pol.}_{SM}|^2$  can be written  with the help of 
\begin{eqnarray}
a_{SM}^{gg} &=& C_{SM}^{gg} \left [ 1  + 2 \beta^2 (1 -z^2) - (2 - 2 z^2 + z^4) \beta^4 \right ]\,,
\nonumber \\
c_{SM}^{gg} &=&  C_{SM}^{gg} \left [ 1 - 2 \beta^2 + (2 - 2 z^2 + z^4) \beta^4  \right ]\,,
\nonumber \\
e_{SM}^{gg} &=&  C_{SM}^{gg} \left [  - \frac{4}{s} \beta^2 ( 1+ \beta z) (1 - z^2) \right ]\,,
\nonumber \\
f_{SM}^{gg} &=& C_{SM}^{gg} \left [  - \frac{4}{s} \beta^2 (1 - \beta z) (1 - z^2) \right ] \,,
\end{eqnarray}
where $C_{SM}^{qq} = 32 \pi^2 \alpha_s^2 \frac{(N^2 (1 + \beta^2 z^2) -2)}{(1 - \beta^2 z^2)^2}  \frac{(N^2 -1)}{ N} $. These expressions can be also found in Secs. 2.1.1. and 2.1.2 of~\cite{spinSM}. 

\subsection{{ New Physics Models}}

\subsubsection{{ Axigluon }}

The relevant interaction part of the Lagrangian describing new physics in the $t\bar{t}$ production due to s-channel axigluon ${G'}$ exchange is given by\,,
\begin{equation}
  {\cal L}^{\rm int.}_{G'}  = -  \bar q  (g_V^q - g_A^q \gamma_5) {\DS{G'}}  q   - \bar t  ( g_V^t - g_A^t \gamma_5) {\DS{ G'}}  t  \,,
\label{Lag}
\end{equation} 
The new $q\bar q \to t\bar t$ amplitude is then
\begin{equation}
i \mathcal M_{G'}= i T^a_{ij} T^a_{kl}  \frac{1}{s - m_{G}^2 + i m_G \Gamma_G} \bar v^i (p_2) \gamma^\mu (g_V^q + g_A^q \gamma_5)  u^j(p_1) \bar u^k( k_1,s_t) \gamma_\mu (g_V^t + g_A^t \gamma_5) v^l (k_2, s_{\bar t})\,,
\label{Mag}
\end{equation}
where $m_G$ is the axigluon mass and $\Gamma_G$ is the manually introduced axigluon width. The resulting unpolarized $q\bar{q} \to t\bar t$ cross section is of the form~\cite{axig2} 
\begin{eqnarray}
\frac{d \sigma_{SM+G'}^{q\bar{q}, \rm unpol.}}{d t} & =& \frac{d \sigma_{SM}^{q\bar{q}, \rm unpol.}}{d t} 
+ \frac{2 \pi}{9 s^2} \alpha_s^2 \Bigg \{ 
\frac{s( s- m_{G}^2)}{(s- m_{G}^2)^2 + \Gamma_{G}^2 m_{G}^2 }  2 \left ( \bar g_V^q \bar g_V^t (1 + \beta^2 z^2 + 4 m_t^2/s ) + 2 \bar g_A^q \bar g_A^t \beta z \right )
\nonumber \\
&+&   \frac{s^2 }{(s- m_{G}^2)^2 + \Gamma_{G}^2 m_{G}^2 } 
\left [   \left ( (\bar g_V^q)^2 + (\bar g_A^g)^2  \right ) \left (  (\bar g_V^t)^2 ( 1 + \beta^2 z^2 + 4 m_t^2/s ) 
\right . \right .  \nonumber \\
&& \left . \left .  + (\bar g_A^t)^2 ( 1 + \beta^2 z^2 - 4 m_t^2/s ) + 8 \beta z \bar g_V^q \bar g_V^t \bar g_A^q \bar g_A^t \right ) \right ]  \Bigg \}\,,
\label{sigma-ag}
\end{eqnarray}
where now $\bar g_{V,A}^{q,t} = g_{V,A}^{q,t}/g_s$. 

The polarized parts are given (for $g_V^q = g_V^t = 0$) by the following expressions, eq. (\ref{eq:Mpol}):
\begin{eqnarray}
a_{SM*G'} &=& C_{SM*G'} [ 4 \beta z ]\,,
\nonumber \\
e_{SM*G'} &=& C_{SM*G'} \left [ \frac{ 16 m_t^2}{s^2} \right ]\,,
\nonumber \\
f_{SM*G'} &=& C_{SM*G'} \left [ \frac{ -16 m_t^2}{s^2} \right ]\,,
\end{eqnarray}
and 
\begin{eqnarray}
a_{G'} &=& C_{G'} [1 - 4m_t^2/s + \beta^2 z^2] \,,
\nonumber \\
c_{G'} &=&  C_{G'} [-1 + 4m_t^2/s + \beta^2 z^2] \,,
\nonumber \\
e_{G'} &=&  C_{G'} \left [ \frac{4}{s} (1 - 4m_t^2/s + \beta z) \right ] \,,
\nonumber \\
f_{G'} &=& C_{G'} \left [ \frac{4}{s} (-1 + 4m_t^2/s + \beta z) \right ] \,,
\end{eqnarray}
where 
\begin{eqnarray}
C_{SM*G'} &=& 16 \pi^2 \alpha_s^2 (N^2 -1) \frac{g_A^q \bar g_A^t \, s (s - m_{G}^2)}{ (s - m_{G}^2)^2 + \Gamma_{G}^2 m_{G}^2 } \, , \nonumber\\ 
C_{G'} &=& 16 \pi^2 \alpha_s^2 (N^2 -1) \frac{(\bar g_A^q)^2 (\bar g_A^t)^2 \, s^2 }{(s- m_{G}^2)^2 + \Gamma_{G}^2 m_{G}^2 }\,.
\end{eqnarray}
Other coefficients in  (\ref{eq:Mpol}) are vanishing. 

\subsubsection{{$Z'$ and $W'$}}

The interactions of the heavy weak bosons $Z'$ and $W'$ contributing in the t-channel of the $t\bar{t}$ production are given by the following Lagrangians:
\begin{equation}
  {\cal L}^{\rm int.}_{Z'} =  - \bar u \gamma_{\mu} ( f_L^{Z'} P_L  + f_R^{Z'} P_R ) t Z'   +   \rm  h.c.\, ,
\label{LZ'}
\end{equation}
\begin{equation}
  {\cal L}^{\rm int.}_{W'} =  - \bar d \gamma_{\mu} ( f_L^{W'} P_L  + f_R^{W'} P_R ) t W'  + \rm  h.c.\,.
\label{LZ'}
\end{equation}
The relevant amplitude induced by the new physics contributions is
\begin{eqnarray}
i \mathcal M_{Z'} = i \bar u(k_1,s_t) \gamma^{\mu} ( f_L^{Z'} P_L  + f_R^{Z'} P_R ) u(p_1)  \left ( -\frac{g^{\mu\nu}}{t - m_{Z'}^2} + \frac{t^{\mu} t^{\nu}}{m_{Z'}^2 ( t - m_{Z'}^2 )} \right ) 
\bar v (p_2) \gamma^{\nu} ( f_L^{Z'} P_L  + f_R^{Z'} P_R ) v( k_2,s_{\bar t} )
\nonumber \\
\end{eqnarray}
where $t^\mu = (p_1-k_1)^\mu$\,, and similarly for the interaction with the $W'$, by the exchange $Z' \to W'$ and $u \to d$. 
Then
\begin{eqnarray}
\frac{d \sigma^{u\bar{u}, \rm unpol.}_{SM+Z'}}{d t} &=& \frac{d \sigma_{SM}^{u\bar{u}, \rm unpol.}}{d t}  + \frac{\alpha_s}{9 s^3} \frac{(f_L^{Z'})^2 + (f_R^{Z'})^{2}}{t - m_{Z'}^2}
 \{ 2 u_t^2 + 2 s m_t^2 +\frac{m_t^2}{m_{Z'}^2} ( t_t^2 + m_t^2 s)\}
\nonumber \\
&+& \frac{1}{16 \pi s^2}
\frac{1}{(t- m_{Z'}^2)^2}  \left \{ \left ((f_L^{Z'})^{4} + (f_R^{Z'})^{4} \right )  u_t^2 + 2 (f_L^{Z'})^{2} (f_R^{Z'})^{2} s (s-2 m_t^2)  
\right . 
\nonumber \\
&& \left . +\frac{m_t^4}{4 m_{Z'}^4}\left ((f_L^{Z'})^{2} + (f_R^{Z'})^{2} \right )^2 ( t_t^2 + 4 s m_{Z'}^2) \right \}\,,
\label{sigma-WZ}
\end{eqnarray}
where $u_t = s ( 1 + \beta z)/2, t_t = s ( 1 - \beta z)/2$.

The polarized case we consider only the right-handed couplings and the non-vanishing spin coefficients in (\ref{eq:Mpol}) are
\begin{eqnarray}
a_{SM*Z'} &=& C_{SM*Z'} \left [ 4 m_t^2 + s ( 1 + \beta z)^2  + \frac{m_t^2}{2 m_{Z'}^2}(4 m_t^2 + s ( 1 - \beta z)^2 )   \right ]\,,
\nonumber \\
c_{SM*Z'} &=& C_{SM*Z'} \left [ - 4 m_t^2 + s ( 1 - \beta^2 z^2)  - \frac{m_t^2}{ 2 m_{Z'}^2}(4 m_t^2 - s ( 1 - \beta^2 z^2 ))  \right ]\,,
\nonumber \\
e_{SM*Z'} &=& C_{SM*Z'} \left [ \frac{4}{s} \left  (  2 m_t^2 - s ( 1 + \beta z)  - \frac{m_t^2}{ 2 m_{Z'}^2} ( 2 m_t^2 +s (1 + \beta z))\right  ) \right ]\,,
\nonumber \\
f_{SM*Z'} &=& C_{SM*Z'} \left [ \frac{-4}{s} \left (  2 m_t^2 + s ( 1 - \beta z)  - \frac{m_t^2}{ 2 m_{Z'}^2} ( 2 m_t^2 -s (1 - \beta z))  \right  ) \right ]\,,
\nonumber \\
(b_1^t)_{SM*Z'} &=& C_{SM*Z'} \left [ - 2 m_t  \left  (  1 + \beta z +  \frac{m_t^2}{ 2 m_{Z'}^2} ( 3 - \beta z) \right  ) \right ]\,,
\nonumber \\
(b_2^t)_{SM*Z'} &=& C_{SM*Z'} \left [ - 2 m_t  \left  (  3 + \beta z +  \frac{m_t^2}{ 2 m_{Z'}^2} ( 1 - \beta z) \right  ) \right ]\,,
\nonumber \\
\end{eqnarray}
and 
\begin{eqnarray}
a_{Z'} &=& C_{Z'} \left [  s ( 1 + \beta z)^2 +  4 \frac{m_t^4}{ m_{Z'}^2} + \frac{m_t^4}{ 4 m_{Z'}^4} s (1 - \beta z )^2  \right ]\,,
\nonumber \\
c_{Z'} &=& C_{Z'} \left [  \frac{m_t^2}{m_{Z'}^2} \left ( s ( 1 - \beta^2 z^2) -  4 m_t^2 \right ) \right ]\,,
\nonumber \\
e_{Z'} &=& C_{Z'} \left [ \frac{4}{s}   \frac{m_t^2}{m_{Z'}^2} \left (  2 m_t^2 - s ( 1 + \beta z)  - \frac{m_t^4}{ m_{Z'}^2} \right ) \right ]\,,
\nonumber \\
f_{Z'} &=& C_{Z'} \left [ \frac{4}{s}   \frac{m_t^2}{m_{Z'}^2} \left (  2 m_t^2 - s ( 1 - \beta z)  - 4  m_{Z'}^2 \right ) \right ]\,,
\nonumber \\
(b_1^t)_{Z'} &=& C_{Z'} \left [ - 2 m_t   \frac{m_t^2}{ m_{Z'}^2}  \left ( ( 1 + \beta z) +  \frac{m_t^2}{ 2 m_{Z'}^2} ( 1 - \beta z) \right ) \right ] \,,
\nonumber \\
(b_2^t)_{Z'} &=& C_{Z'} \left [ - 4 m_t  \left ( ( 1 + \beta z) +  \frac{m_t^2}{ 2 m_{Z'}^2} ( 1 - \beta z) \right ) \right ] \,,
\nonumber \\
\end{eqnarray}
with 
\begin{eqnarray}
C_{SM*Z'} = 8 \pi \alpha_s \frac{f_R^2}{  t- m_{Z'}^2 } \frac{N^2 -1}{2} 
\,, \qquad 
C_{Z'} = f_R^4 \frac{s}{( m_{Z'}^2 -t)^2} N^2 \,.
\end{eqnarray}
The $W'$ case is again given by the same expressions by substituting $Z' \to W'$ everywhere.

\subsubsection{{Scalar isodoublet }}

The interactions of the weak doublet scalar $\Phi \sim (1,2)_{1/2}$ with quarks are given by~\cite{doublet, AguilarSaavedra:2011zy}
\begin{equation}
  {\cal L}^{\rm int.}_{\Phi} =  - y_{ij}^{u}   \bar q _{Li}  u_{Rj} \Phi  -   y_{ij} ^{d}  \bar q _{Li}  d_{Rj}  \tilde \Phi + \rm h.c.\,,
\label{Ld}
\end{equation}
where $\tilde \Phi = i \tau_2 \Phi^*$. The amplitude coming from the exchange of the neutral isodoublet component $\phi^0$ is then (with $y_{ij}^{u} = y_{ij}$)
\begin{eqnarray}
i \mathcal M_{\phi}& =&  -i \frac{|y_{31}|^{2} }{8 (t-m_\phi^2) } \bar u^a (k_1,s_t) \gamma^\mu  (1-\gamma_5) v^b ( k_2,s_{\bar t} ) 
\bar v^b (p_2) \gamma_ {\mu}  (1+\gamma_5)  u^a(p_1)\,,
  \label{M-2}
\end{eqnarray}
where $m_\phi$ is the $\phi^0$ mass, resulting in the cross section
\begin{eqnarray}
\frac{d \sigma^{u\bar{u}, \rm unpol.}_{SM+\phi}}{d t} & =& \frac{d \sigma_{SM}^{u\bar{u}, \rm unpol.}}{d t}    - \frac{\alpha_s}{9 } \frac{| y_{13}|^2}{s^3} \frac{m_t^2 s + (m_t^2-t)^2}{m_{\phi}^2 -t} 
 +\frac{ | y_{13}|^4}{64 \pi} \frac{1}{s^2}  \frac{(m_t^2- t)^2 }{(m_{\phi}^2-t)^2}.  
\label{sigma-doublet}
\end{eqnarray}

The coefficients in the polarized cross section are
\begin{eqnarray}
a_{SM*\phi} &=& C_{SM*\phi} \left [ t^2  + m_t^2 (s - 2 t ) + m_t^4 \right ]\,,
\nonumber \\
c_{SM*\phi} &=& C_{SM*\phi} \left [ t u - m_t^4 \right ]\,,
\nonumber \\
e_{SM*\phi} &=& C_{SM*\phi} \left [ (-2)(s + t) \right ]\,,
\nonumber \\
f_{SM*\phi} &=& C_{SM*\phi} \left [ 2 t \right ]\,,
\nonumber \\ 
(b_1^t)_{SM*\phi} &=& C_{SM*\phi}  \left [ m_t (u + 2 t -3 m_t^2 ) \right ]\,,
\nonumber \\ 
(b_2^t)_{SM*\phi} &=& C_{SM*\phi}  \left [ m_t ( t - m_t^2) \right ] \,,
\end{eqnarray}
and
\begin{eqnarray}
a_{\phi} &=& C_{\phi} \left [ (m_t^2 -t)^2 \right ]\,,
\nonumber \\
e_{\phi} &=& C_{\phi} \left [ -4 m_t^2 \right ]\,,
\nonumber \\
(b_1^t)_{\phi} &=& C_{\phi} \left [ 2 m_t( t - m_t^2 )\right ]\,,
\end{eqnarray}
where 
\begin{eqnarray}
C_{SM*\phi} = \frac{16 \pi \alpha_s}{s} \frac{ | y_{13}|^2}{ ( t- m_{\phi}^2 )} \frac{N^2 -1}{2}
\,, \qquad 
C_{\phi} = \frac{| y_{13}|^4}{( m_{\phi}^2 -t)^2} N^2 \,.
\end{eqnarray}
The rest of the coefficients are vanishing. 

\subsubsection{{ Scalar color triplet and sextet}}

The relevant interaction Lagrangians in this case are  given by
\begin{equation}
  {\cal L}^{\rm int.}_{\Delta} = -  {g_{(\Delta)ij}}_{}  \epsilon_{abc} \bar u_{R,i}^a ( u_{R,j}^{b})^C \Delta^C + \rm h.c.\, ,
\label{L3}
\end{equation} 
with $\Delta^C$ being a $(3,1,-4/3)$ state, and  
\begin{equation}
  {\cal L}^{\rm int.}_{\Sigma} = -  {{g_{(\Sigma)ij}}_{}}{}   \ ( \bar u_{R,i}^a   ( u_{R,j}^{b} )^C + \bar u_{R,i}^b   ( u_{R,j}^{a} )^C   ) \Sigma^{ab\dag} + \rm h.c.\,,
\label{L6}
\end{equation} 
with $\Sigma^{ab}$ being a $(\bar 6 ,1, 4/3)$ state. 
For the color triplet amplitude one finds
\begin{eqnarray}
i \mathcal M_{\Delta} &=& 
  i \frac{ |{g_{(\Delta)13}}_{}|^2}{8} \frac{1}{u - m_{\Delta}^2}[ \bar u^a (k_1,s_t) \gamma^{\mu}(1+\gamma_5)  v^a ( k_2,s_{\bar t} )  \bar v^b(p_2) \gamma_{\mu} (1+\gamma_5) u^b (p_1) 
\nonumber \\
&&   -  \bar u^a (k_1,s_t) \gamma^{\mu}(1+\gamma_5)  v^b ( k_2,s_{\bar t} )  \bar v^b(p_2) \gamma_{\mu} (1+\gamma_5) u^a (p_1) ]\,,
  \label{M-3}
\end{eqnarray}
while for the color sextet case we have 
\begin{eqnarray}
i M_{\Sigma} &=& 
  i \frac{ |g_{(\Sigma)13}|^2}{8} \frac{1}{u - m_{\Sigma}^2}[ \bar u^a (k_1,s_t) \gamma^{\mu}(1+\gamma_5)  v^a ( k_2,s_{\bar t} )  \bar v^b(p_2) \gamma_{\mu} (1+\gamma_5) u^b (p_1) 
\nonumber \\
&&   +  \bar u^a (k_1,s_t) \gamma^{\mu}(1+\gamma_5)  v^b ( k_2,s_{\bar t} )  \bar v^b(p_2) \gamma_{\mu} (1+\gamma_5) u^a (p_1) ]\,.
  \label{M-6}
\end{eqnarray}
The cross section for the color triplet scalar is then given by 
\begin{eqnarray}
\frac{d \sigma_{SM+\Delta}^{u\bar{u}, \rm unpol.}}{d t} & =& \frac{d \sigma_{SM}^{u\bar{u}, \rm unpol.}}{d t} - \frac{\alpha_s}{9} \frac{| g_{(\Delta)13}|^2}{s^3} \frac{m_t^2 s + (m_t^2-u)^2}{m_{\Delta}^2 -u} 
 +\frac{ | g_{(\Delta)13}|^4}{48 \pi} \frac{1}{s^2}  \frac{(m_t^2- u)^2 }{(m_{\Delta}^2 - u)^2 }\,.  
\label{sigma-3}
\end{eqnarray}
In the case of the color sextet model one finds instead 
\begin{eqnarray}
\frac{d \sigma_{SM+\Sigma}^{u\bar{u}, \rm unpol.}}{d t} & =& \frac{d \sigma_{SM}^{u\bar{u}, \rm unpol.}}{d t} + \frac{\alpha_s}{9} \frac{| g_{(\Sigma)13}|^2}{s^3} \frac{m_t^2 s + (m_t^2-u)^2}{m_{\Sigma}^2 -u} 
 +\frac{ | g_{(\Sigma)13}|^4}{24 \pi} \frac{1}{s^2}  \frac{(m_t^2- u)^2 }{(m_{\Sigma}^2 -u)^2}\,.  
\label{sigma-6}
\end{eqnarray}

The nonvanishing polarization coefficients for these models are: 
\begin{eqnarray}
a_{SM*\Delta/\Sigma} &=& C_{SM*\Delta/\Sigma} \left [  u^2  + m_t^2 (s - 2 u ) + m_t^4 \right ]\,,
\nonumber \\
c_{SM*\Delta/\Sigma} &=& C_{SM*\Delta/\Sigma} \left [  t u - m_t^4 \right ]\,,
\nonumber \\
e_{SM*\Delta/\Sigma} &=& C_{SM*\Delta/\Sigma} \left [ 2 u \right ]\,,
\nonumber \\
f_{SM*\Delta/\Sigma} &=& C_{SM*\Delta/\Sigma} \left [ -2( s + u)\right ]\,,
\nonumber \\ 
(b_1^t)_{SM*\Delta/\Sigma} &=& C_{SM*\Delta/\Sigma}  \left [ m_t ( u - m_t^2) \right ]\,,
\nonumber \\ 
(b_2^t)_{SM*\Delta/\Sigma} &=& C_{SM*\Delta/\Sigma}  \left [ \pm m_t (3 m_t^2 -t - 2 u)\right ]\,,
\end{eqnarray}
and
\begin{eqnarray}
a_{\Delta/\Sigma} &=& C_{\Delta/\Sigma} \left [ (m_t^2 -u)^2 \right ]\,,
\nonumber \\
e_{\Delta/\Sigma} &=& C_{\Delta/\Sigma} \left [ -4 m_t^2 \right ] \,,
\nonumber \\
(b_2^t)_{\Delta/\Sigma} &=& \pm C_{\Delta/\Sigma} \left [ 2 m_t( m_t^2 -u ) \right ]\,,
\end{eqnarray}
where
\begin{eqnarray}
C_{SM*\Delta/\Sigma} = \pm \frac{8 \pi \alpha_s}{s} \frac{ | (g_{(\Delta/\Sigma)13}|^2}{ ( u- m_{\phi}^2 )} (N^2 -1)
\,, \qquad 
C_{\Delta/\Sigma} = \frac{| g_{(\Delta/\Sigma)13}|^4}{( m_{\phi}^2 -u)^2} [ 2 N (N \mp 1) ]\,.
\end{eqnarray}

\end{document}